\documentclass[12pt]{article}
\usepackage{amsfonts,amssymb}
\normalsize
\newcounter{saveequation} 

\makeatletter
\@addtoreset{equation}{section}
\makeatother
\newcounter{k}

\newcommand{\ol}[1]{\overline{#1}}

\def \ket{\rangle}

\def \bra{\langle}

\newcommand{\RR}{\ensuremath{\mathbb{R}}}
\newcommand{\NN}{\ensuremath{\mathbb{N}}}
\newcommand{\ZZ}{\ensuremath{\mathbb{Z}}}
\newcommand{\CC}{\ensuremath{\mathbb{C}}}

\newcommand{\VV}{\ensuremath{\mathcal{V}}}

\newcommand{\HHH}{\ensuremath{\mathcal{H}}}
\newcommand\p{^\prime\!}
\newcommand{\g}{\mathfrak{g}}
\newcommand{\h}{\mathfrak{h}}
\newcommand{\ab}{\mathfrak{a}}
\def \beq{\begin{equation}} \def\eeq{\end{equation}}
\def \beqn{\begin{eqnarray}} \def\eeqn{\end{eqnarray}}
\def\el#1{\label{#1}}
\def\eq#1{(\ref{#1})}
\begin{document}  
\thispagestyle{empty}
\begin{flushright}  
                     hep-th/0304234

KCL-MTH-03-03
\end{flushright}  
\vskip 2cm    

\begin{center}  
{\huge   Current-Current Deformations of}
\end{center}\begin{center}{\huge Conformal Field Theories, and WZW Models}
\vspace*{5mm} \vspace*{1cm}   
\end{center}  
\vspace*{5mm} \noindent  
\vskip 0.5cm  
\centerline{\bf Stefan F\"orste$^a$ and Daniel Roggenkamp$^{a,b}$}
\vskip 1cm
\centerline{\em ${}^a$ Physikalisches Institut, Universit\"at Bonn}
\centerline{\em Nussallee 12, D-53115 Bonn, Germany}

\vskip .3cm

\centerline{\em ${}^b$ Department of Mathematics,  
King's College London}  
\centerline{\em Strand, London WC2R 2LS, United Kingdom}

\vskip2cm
  
\centerline{\bf Abstract}  
\vskip .3cm

Moduli spaces of conformal field theories corresponding 
to current-current deformations are discussed. For WZW models, 
CFT and sigma model considerations are compared. It is shown 
that current-current deformed WZW models have WZW-like sigma model 
descriptions with non-bi-invariant metrics, additional $B$-fields
and a non-trivial dilaton.

\vskip .3cm

\newpage
\section{Introduction}
Moduli spaces of conformal field theories are important objects
in the study of two dimensional quantum field theories because
they describe critical subspaces in the space of coupling constants.
In string theories, whose small coupling limits are described by conformal field theories,
these moduli spaces arise as parameter spaces of string vacua. The 
understanding of moduli spaces of conformal field theories is thus a very important issue for
string theory. 

Although conformal field theories are quite well understood, there are
only few examples of explicitly known moduli spaces at present, most of which
correspond to free field theories as e.g.\ toroidal conformal field theories 
or orbifolds thereof. 

The reason for this is that a good conceptual understanding of
deformations of conformal field theories beyond conformal perturbation
theory is still lacking. Perturbation theory is usually technically involved, 
at least when one wants to obtain higher order contributions, and hence
is only applicable to study CFT moduli spaces in small neighbourhoods of 
explicitly known models. In particular it is in general not possible 
to obtain global information about the moduli spaces from perturbation
theory, except in situations, where symmetry as e.g.\ supersymmetry is
preserved by the corresponding deformations.

In this article, special deformations of conformal field theories, namely those
generated by perturbations with products of holomorphic and
antiholomorphic currents are studied. 
As was shown in \cite{Chaudhuri:1988qb}, such deformations preserve
the algebras  
of the perturbing fields. Indeed, this gives enough structure to determine 
global properties of the subspaces of moduli spaces corresponding to this kind
of deformations. 

Important examples of non-free conformal field theories 
admitting current-current deformations are WZW models (see
e.g.\ \cite{Gepner:1986wi}). The corresponding moduli spaces for WZW models 
associated to compact semi-simple Lie groups, will be discussed in detail.

Since WZW models have descriptions as sigma models on Lie groups,
it is a natural question if there are such descriptions for all
conformal field theories from these moduli spaces. 
In fact, families of sigma models containing WZW models at special
points have been discussed by many 
authors (e.g.\ in
\cite{Hassan:1992gi,Giveon:1993ph,Sfetsos:1993ka,Kiritsis:1993ju,Gershon:1993wu}). 
In particular one-parameter families of sigma models 
containing the ${\rm SU}(2)$-WZW models have been studied very explicitly by
Giveon and Kiritsis  \cite{Giveon:1993ph}, who also compared them
to the families of current-current deformed ${\rm SU}(2)$-WZW models which were
described in \cite{Yang:1988bi}. 
Ideas about a generalization of these 
considerations to arbitrary WZW models have also been presented in
\cite{Giveon:1993ph,Kiritsis:1993ju,Kiritsis:1991zt}.

In this article, WZW-like sigma model representations 
of current-current deformed WZW models will be
explicitly constructed.  
These are sigma models with the same target space as the
``undeformed'' WZW model in the family, but 
with different (in general not bi-invariant) metrics and additional $B$-fields. 

Thus, we obtain explicit descriptions of the moduli spaces of
current-current deformed WZW models associated to compact semi-simple Lie groups
in terms of conformal field theories as well as in terms of sigma models. 

In section two, exactly marginal current-current deformations of conformal
field theories are discussed.
We start from the facts obtained in \cite{Chaudhuri:1988qb},
that perturbations of conformal field theories
with products of holomorphic and antiholomorphic currents are exactly 
marginal iff the holomorphic as well as the antiholomorphic currents 
belong to commutative current algebras, and that in this case
these holomorphic and antiholomorphic current algebras are preserved
under the deformations. 
This can be used to reduce the problem of studying finite
current-current deformations 
to first order deformation theory, which is carried out in
\ref{deformationtheory},  
and from which it follows that the 
effect of these deformations on the CFT structures is completely captured by
pseudo orthogonal transformations of their charge lattices with respect 
to the preserved commutative current algebras. 
The corresponding deformation spaces can thus be described by
\beq
{\cal D}\simeq{\rm O}(d,{\bar{d}})/{\rm O}(d)\times{\rm O}(\bar{d})\,.
\eeq
This generalizes the deformation results for toroidal conformal field 
theories \cite{Dijkgraaf:jt}.
The corresponding moduli spaces are obtained from the deformation spaces
by taking the quotients with respect to ``duality groups''.

In section three we discuss an important class of examples, namely WZW models
corresponding to compact semi-simple Lie-groups.
The general results from section two are compared to a realization of
deformed WZW models obtained from a representation of WZW models as 
orbifolds of products of generalized parafermionic 
and toroidal models given in \cite{Gepner:1987sm}. 

In section four we discuss various aspects of exactly marginal
deformations of WZW models from a sigma model perspective. This
approach is best suited for a semiclassical treatment and in that
sense less powerful than the algebraic one. However, it can illustrate
the results and provide a picture for the class of deformed models. 
Exactly marginal deformations of WZW models from the sigma model
perspective have been discussed in the past mainly for rank one
groups \cite{Hassan:1992gi,Giveon:1993ph,Sfetsos:1993ka} and for models
where coordinates can be chosen 
such that the relevant set of chiral and anti-chiral currents follows
manifestly from the equations of motion \cite{Henningson:1992rn}.

Mimicking the orbifold realization of deformed WZW models described
in section three, we will consider an orbifold of a direct product
consisting of a vectorially gauged WZW model and a 
{\it d}-dimensional torus model, where {\it d} is the rank of the group. 
Since a sigma model is not very well designed to accommodate
orbifolds, we perform an 
axial-vector duality (generalized T-duality) to obtain a dual
description without an additional orbifold action. To this end, we
first implement the orbifold by gauging in addition an axial symmetry
of the WZW model combined with shifts in the torus factor. We force
the corresponding gauge bundle to be flat but choose the zero modes of
the corresponding Lagrange multiplier such that the gauge bundle is
twisted in a non-trivial way. It turns out that integrating out the
gauge bundle instead of the Lagrange multiplier provides a sigma model
without an additional orbifold action. The result is a ``WZW-like''
model, i.e.\ a sigma model with Lie group as target space, and
an action given by a WZW action in which the bi-invariant metric is
replaced by a more general bilinear form which is neither bi-invariant
nor necessarily symmetric.

The same sigma models can be obtained as coset models of  
products of the original WZW models and {\it d}-dimensional torus models
with gauge group ${\rm U}(1)^d$, embedded into both factors.

All the sigma model manipulations described so far are 
carried out at a classical level. Quantum
mechanically one should replace the procedure of solving equations of
motion by performing Gaussian integrals. These in general provide
functional determinants, which in turn generate a non-trivial
dilaton. In a pragmatic approach this can be computed by imposing
conformal invariance, i.e.\ requiring vanishing beta functions. 
We will use a more elegant way consisting of a comparison of the
Hamiltonian of the model and a generalized Laplacian, which depends 
on the dilaton \cite{Tseytlin:1993my}. 

Finally, we illustrate some of the results in the 
example of the deformed ${\rm SU}(2)$-WZW model.

We should mention that many of the employed techniques
can be found in the literature. The corresponding references will be
given in the text.
\section{Current-Current Deformations of CFTs}
\label{ccdef}

Although not proven in general, it is widely believed that all deformations
of conformal field theories are generated by perturbations of the
theories with marginal fields, i.e.\ fields ${\cal O}_i$ with conformal
weights $h({\cal O}_i)=1=\ol{h}({\cal O}_i)$. 

The perturbed correlation functions on a conformal 
surface $\Sigma$ of a combination of 
operators $X(p_1,\ldots,p_k)$, $p_i\in\Sigma$ are defined to be
\beq\el{deformation}
\bra X(p_1,\ldots,p_k)\ket_\Sigma^\lambda:=\bra X(p_1,\ldots,p_k)
\exp{\left(\sum_i\lambda_i\int_\Sigma {\cal O}_i d\mu_\Sigma\right)}
\ket_\Sigma\,,  
\eeq
where the integrals have to be regularized due to the appearance of
singularities. If the perturbed correlation functions
define a quantum field theory, which is a fixed point of the
renormalization group flow, it is again a conformal field theory. This
however is not the case in general. Indeed, 
preservation of conformal invariance by perturbations gives non linear
restrictions on the fields, which generate it (see
e.g.\ \cite{Dijkgraaf:jt}).  
This means that the set of
exactly marginal fields, i.e.\ those which generate deformations of conformal 
field theories are not vector spaces in general. In particular the
deformation spaces 
of conformal field theories need not necessarily be manifolds but may have 
singularities. (For more details on conformal deformation theory see
e.g.\ \cite{Dijkgraaf:jt,Kutasov:1988xb,Ranganathan:1992nb,Ranganathan:1993vj}).  

In the following, deformations of conformal field theories generated
by a special class of marginal  
fields, namely products of holomorphic and antiholomorphic currents
will be discussed.  
These are simple enough to give a global description of 
the deformation spaces corresponding to them 
and to express the data of deformed CFTs explicitly in terms of the data of 
the undeformed ones.

We consider conformal field theories, whose holomorphic 
and antiholomorphic $W$-algebras contain current algebras $\hat{\g}_k$, 
$\hat{\ol{\g}}_{\ol{k}}$ corresponding to Lie algebras $\g$ and $\ol{\g}$ and
$k,\ol{k}\in\NN$, i.e.\ for every $j,j\p \in \g$ there exist
holomorphic fields $j(z),j\p(z)$ of  
conformal weight $h=1$ in the theory, such that
\beqn\el{currentalg}
j(z)j\p(w)={k K_{\g}(j,j\p)\over (z-w)^2}+{[j,j\p](w)\over (z-w)}+{\rm reg}\\
T(z)j(w)={j(w)\over (z-w)^2}+{\partial j(w)\over (z-w)}+{\rm reg}\,,\nonumber
\eeqn
where $K_{\g}(.,.)$ is a bi-invariant scalar product on $\g$ and
$[.,.]$ its Lie bracket.  
The holomorphic energy momentum tensor $T(z)$ can be written as
\beq
T(z)=T_0(z)+{1\over 2(k+\check{g})}
\sum_{\alpha,\beta} \kappa^g_{\alpha\beta}:\!j^\alpha(z)j^\beta(z)\!:\,,
\quad \kappa^\g:=K^{-1}_\g\,,
\eeq
with $\check{g}$ the dual Coxeter number of $\g$,
$(j^\alpha)_\alpha$ a basis of $\g$ and $T_0(z)j(w)={\rm reg}$.
The same holds for the antiholomorphic current algebra with $(\g,k)$ replaced
by $(\bar{\g},\bar{k})$.

In particular, there is a subspace of the CFT Hilbert space isomorphic to 
$\g\otimes\ol{\g}$ of marginal fields. However, not all of these fields
are exactly marginal. As was shown in \cite{Chaudhuri:1988qb}, such
fields are exactly marginal if and only if, under the isomorphism
above, they correspond to elements of $\ab\otimes\ol{\ab}$, 
for any abelian subalgebras $\ab\subset\g$,
$\ol{\ab}\subset\ol{\g}$. Moreover deformations 
generated by these exactly marginal fields preserve the current
algebras $\hat{\ab}$, $\hat{\ol{\ab}}$ corresponding to $\ab$, $\ol{\ab}$.

Hence, every pair of abelian $\hat{u}(1)^d\simeq \ab\subset \g$, 
$\hat{\ol{u}}(1)^{\ol{d}}\simeq\ol{\ab}\subset\ol{\g}$ gives rise to a
family of  
conformal field theories with current algebras $\hat{\ab}$, $\hat{\ol{\ab}}$.
All these families meet in the original model. 
Since conformal field theories are invariant under automorphisms of their
holomorphic and antiholomorphic $W$-algebras, pairs of abelian
subalgebras which are identified
under such automorphisms give rise to equivalent deformations.

We assume in the following that the conformal field theory is unitary,
and its Hilbert space  
decomposes into tensor products of
irreducible highest weight representations $\VV_Q$, $\ol{\VV}_{\ol{Q}}$ of 
$\hat{\ab}$, $\hat{\ol{\ab}}$, which are characterized
by their charges $Q\in \ab^*$, $\ol{Q}\in\ol{\ab}^*$, and whose lowest
conformal 
weights are given by $h_Q={1\over 2}\kappa(Q,Q)$, $\ol{h}={1\over
  2}\ol{\kappa}(\ol{Q},\ol{Q})$, as can 
be read off from \eq{currentalg}\footnote{From now on $\kappa={1\over
    k}\kappa^\g|_\ab$, $\ol{\kappa}={1\over
    k}\kappa^{\ol{\g}}|_{\ol{\ab}}$.}: 
\beq\el{decomp}
\HHH\simeq
\bigoplus_{(Q,\ol{Q})\in\Lambda}\HHH_{Q\ol{Q}}\otimes\VV_Q\otimes
\ol{\VV}_{\ol{Q}}\,.   
\eeq
The set of charges $\Lambda\subset \ab^*\times \ol{\ab}^*$ forms a lattice
equipped with bilinear pairing  
$\bra.,.\ket:=\kappa-\ol{\kappa}$. 

As is shown in \ref{deformationtheory}, deformations corresponding to
pairs $(\ab,\ol{\ab})$ only affect the representation 
theory of the $W$-algebras $\hat{\ab}$, $\hat{\ol{\ab}}$, but not the
OPE-coefficients of $\hat{\ab}\times\hat{\ol{\ab}}$-highest  
weight vectors. More precisely, if one chooses suitable connections on
the bundles of Hilbert spaces over the 
deformation spaces \cite{Ranganathan:1993vj}, the effect of the
deformations on the 
CFT structures 
is completely captured by transformations of the charge
lattices $\Lambda$ in the identity component ${\rm O}(d,\ol{d})_0$ of the
pseudo orthogonal group ${\rm O}(d,\ol{d})$, and all structures
independent of the charges 
are parallel with respect to the chosen connection.

That these ${\rm O}(d,\ol{d})_0$-transformations indeed give rise to 
new modular invariant partition functions and also preserve
locality is easy to see even without any perturbation theory. 
Locality is maintained because of the preservation of 
$\kappa-\ol{\kappa}$ by transformations $O\in {\rm O}(d,\ol{d})$. 
To show the preservation of modular invariance 
we consider the torus partition function depending on modular parameters
$\tau$ and $\ol{\tau}$ ($q=e^{2\pi i\tau}$, $\ol{q}=e^{-2\pi i \ol{\tau}}$)
of the $O(t)$-transformed model along a smooth path
$O:[-1,1]\rightarrow {\rm O}(d,\ol{d})_0$ with $O(0)=1$,  
$\partial_t O(t)\big|_{t=0}=T\in o(d,\ol{d})$
\beqn
Z^{O(t)}(q,\ol{q})&:=&{\rm tr}_\HHH\left(q^{L_0^{O(t)} -{c\over
    12}}\ol{q}^{\ol{L}_0^{O(t)}-{c\over 12}} 
\right)\\
&=&{\rm tr}_\HHH\Big(q^{L_0-{c\over 12}}\ol{q}^{\ol{L}_0-{c\over 12}}
e^{\pi i\Re{(\tau)}\left(O(t)^*-1\right)\bra.,.\ket((j_0,
  \ol{\jmath}_0),(j_0,\ol{\jmath}_0))}\Big. 
\nonumber\\
&&\qquad\qquad\qquad\qquad
\Big. e^{-\pi  \Im{(\tau)}\left(O(t)^*-1\right)\kappa\oplus\ol{\kappa}
  ((j_0,\ol{\jmath}_0),(j_0,\ol{\jmath}_0))} 
\Big)\nonumber\\
&=&
{\rm tr}_\HHH\Big(q^{L_0-{c\over 12}}\ol{q}^{\ol{L}_0-{c\over 12}}
e^{-\pi  \Im{(\tau)}\left(O(t)^*-1\right)\kappa\oplus\ol{\kappa}
  ((j_0,\ol {\jmath}_0),(j_0,\ol{\jmath}_0))} 
\Big)\nonumber
\eeqn
and use the modular transformation properties of the unspecialized characters:
\beqn
\partial_t Z^{O(t)}(q,\ol{q})\big|_{t=0}&=&
\Im{(\tau)}T_{\alpha\ol{\alpha}}{\rm tr}_\HHH\left(j_0^\alpha
\ol{\jmath}_0^{\ol{\alpha}} q^{L_0-{c\over
    12}}\ol{q}^{\ol{L}_0-{c\over 12}}  
\right)\\
&=&
{-\Im{(\tau)}\over 4\pi^2}T_{\alpha\ol{\alpha}}\partial_{z_\alpha}
\partial_{\ol{z}_{\ol{\alpha}}}\big|_{z=\ol{z}=0} 
{\rm tr}_\HHH\left(q^{L_0-{c\over 12}}\ol{q}^{\ol{L}_0-{c\over 12}}
e^{2\pi i \kappa(z,j_0)}e^{-2\pi
  i\ol{\kappa}(\ol{z},\ol{\jmath}_0)}\right) 
\nonumber\\
&=&
{-\Im{(\tau)}\over 4\pi^2}T_{\alpha\ol{\alpha}}\partial_{z_\alpha}
\partial_{\ol{z}_{\ol{\alpha}}}\big|_{z=\ol{z}=0} 
{\rm tr}_\HHH\left(
\tilde{q}^{L_0-{c\over 12}}\tilde{\ol{q}}^{\ol{L}_0-{c\over
    12}}e^{2\pi i \kappa({z\over\tau},j_0)} 
e^{-2\pi i\ol{\kappa}({\ol{z}\over\ol{\tau}},\ol{\jmath}_0)}\right)
\nonumber\\
&=&
\Im{(-{1\over\tau})}
T_{\alpha\ol{\alpha}}{\rm
  tr}_\HHH\left(j_0^\alpha\ol{\jmath}_0^{\ol{\alpha}}
\tilde{q}^{L_0-{c\over 12}}\tilde{\ol{q}}^{\ol{L}_0-{c\over 12}} 
\right)\nonumber\\
&=&
\partial_t Z^{O(t)}(\tilde{q},\tilde{\ol{q}})|_{t=0}\,,\nonumber
\eeqn
where $\tilde{q}=e^{-2\pi i\over\tau}$, $\tilde{\ol{q}}=e^{2\pi i\over
  \ol{\tau}}$.  
This shows that invariance under the 
modular transformation $\tau\mapsto -{1\over\tau}$ 
is preserved under ${\rm O}(d,\ol{d})_0$-transformations of the charge
lattices.  
A similar calculation proves the statement for $\tau\mapsto\tau+1$.

Since charges $(Q,\ol{Q})$ only characterize the
$\hat{\ab}\times\hat{\ol{\ab}}$-modules up to automorphisms 
of the underlying Lie algebras, transformations of $\Lambda$ by 
${\rm O}(d)\times {\rm O}(\ol{d})\subset {\rm O}(d,\ol{d})$ 
leave the conformal field theory invariant.
Thus, the deformation spaces corresponding to pairs $(a,\ol{a})$ are given by
\beq\el{defspace}
{\cal D}_{(a,\ol{a})}\simeq {\rm O}(d,\ol{d})_0/(({\rm O}(d)\times
{\rm O}(\ol{d}))\cap 
{\rm O}(d,\ol{d})_0)\simeq 
{\rm O}(d,\ol{d})/{\rm O}(d)\times {\rm O}(\ol{d})
\,.
\eeq

To get the respective moduli spaces from these deformation spaces, 
one has to identify points describing
equivalent conformal field theories. 
In fact, the conformal field theories are specified by the charge lattices
marked with the Hilbert spaces $\HHH_{(Q,\ol{Q})}$ 
of $\hat{\ab}\times\hat{\ol{\ab}}$-highest weight states 
(with all the structure they carry, as e.g.\ structure of modules 
of the Virasoro algebra etc.) and the coefficients of 
the operator product expansion of these highest weight states. 
Denoting these additional structures by $S$, and the automorphisms of
$\Lambda$  
together with $S$ by ${\rm Aut}(\Lambda,S)$, the components of the moduli
spaces corresponding to $(\ab,\ol{\ab})$ deformations can be written as
\beq\el{defmod}
{\cal M}_{(\ab,\ol{\ab})}
\simeq {\rm Aut}(\Lambda,S)\backslash {\rm O}(d,\ol{d})/{\rm
  O}(d)\times {\rm O}(\ol{d}) 
\,.
\eeq

If the action of ${\rm Aut}(\Lambda,S)$ has fixed points, ${\cal
  M}_{(\ab,\ol{\ab})}$ has singularities and 
the Hilbert space bundle over it has non-trivial monodromies around
  them. More precisely elements in ${\rm Aut}(\Lambda,S)$  
act on the Hilbert space bundles over the deformation spaces, and
  monodromies around fixed points  
are given by the respective actions of the stabilizers.

This gives a very explicit description of the components of moduli
spaces of conformal 
field theories corresponding to current-current deformations. In
particular conformal field  
theories as above come in ${\cal D}_{(\ab,\ol{\ab})}$ families of
explicitly known CFTs, and the 
conformal field theory data at any point in these families can be
easily reconstructed  
from the CFT data at one point. 

A well known example of this kind is the moduli space of toroidal
conformal field theories.  
These models have holomorphic and antiholomorphic $W$-algebras,
each of which contains a $\hat{u}(1)^d$ current-algebra.
They are completely characterized by their charge lattices 
($\HHH_{Q,\ol{Q}}$ are one-dimensional for all charges), 
which for integer spin of the fields, locality and modular 
invariance of the torus partition function have 
to be even, integral, selfdual lattices of signature $(d,d)$ in ${\RR^{d,d}}$
\cite{Casher:1985ra,Narain:1985jj}. 
Hence, $S$ is trivial and ${\rm Aut}(\Lambda,S)\simeq {\rm Aut}(\Lambda)\simeq
{\rm O}(d,d,\ZZ)$, such that 
the moduli spaces ${\cal M}_{d,d}$ corresponding to the current-current 
deformations are isomorphic to the Narain moduli spaces \cite{Narain:1985jj} 
\beq\el{tormod}
{\cal M}^{\rm Narain}_d\simeq {\rm O}(d,d,\ZZ)\backslash {\rm
  O}(d,d)/{\rm O}(d)\times {\rm O}(d)\,, 
\eeq
of even, integral selfdual lattices of 
signature $(d,d)$ in $\RR^{d,d}$. This is the entire moduli space
of toroidal conformal field theories.

A more general class of conformal field theories with 
current algebras are WZW models. These are discussed 
in the next section.
\section{Deformations of WZW Models}
\label{cftdefwzw}
WZW models are conformal field theories associated to Lie groups ${\rm G}$
with bi-invariant  
metrics $\bra .,.\ket$ (see e.g.\ \cite{Gepner:1986wi}).
For simplicity we will only consider compact semi-simple ${\rm G}$ with
bi-invariant metrics 
corresponding to the Killing forms on the respective Lie algebras here.

So, let $k\in\NN$, ${\rm G}$ a semi-simple Lie-group, $\g$ its
Lie-algebra of rank 
$d$ with Killing form $K/k$,  
roots $\Delta$, weights $\Omega$, root lattice ${\rm Q}(\g)$, 
long root lattice ${\rm Q}_l(\g)\subset{\rm Q}(\g)$,
and weight lattice ${\rm P}(\g)$. 
Furthermore denote by $\hat\Omega_k$ the set of integrable weights
of the affine Lie-algebra $\hat{\g}_k$ at level $k$.

The WZW models associated to $({\rm G},K,k)$ have the affine Lie algebras
$\hat{\g}_k$ as holomorphic and  
antiholomorphic $W$-algebras. Its Hilbert spaces decompose into tensor
products of integrable  
highest weight representations $\VV_{\hat{\lambda}}$,
$\hat{\lambda}\in\hat{\Omega}_k$ 
of $\hat{\g}_k$. For simplicity only diagonal WZW models are considered 
in the following, i.e.\ those WZW models whose Hilbert spaces are given by
\beq
\HHH\simeq\bigoplus_{\hat\lambda\in\hat\Omega_k}\VV_{\hat\lambda}
\otimes\ol{\VV}_{\hat\lambda}\,. 
\eeq
Generically, the only marginal fields in WZW models are products of
holomorphic and antiholomorphic currents from the current algebras.  
From the general considerations in section \ref{ccdef} it is clear
that every pair of Cartan subalgebras $\h\subset \g$, $\ol{\h}\subset \g$
gives rise to deformations of the WZW models. 
However, all such pairs lead to equivalent deformations, because all
maximal abelian subalgebras of a semi-simple Lie algebra are pairwise
conjugated\footnote{This is not true for non-semi-simple 
Lie algebras, where one gets more interesting moduli spaces.}. 
Thus the deformation space of current-current-deformed
WZW models is given by
\beq
{\cal D}_{\rm WZW}\simeq {\rm O}(d,d)/{\rm O}(d)\times {\rm O}(d)\,.
\eeq
For a given Cartan subalgebra $\h\subset \g$ the integrable
$\hat{\g}_k$-highest weight representations 
decompose into $\hat{\h}$-highest weight modules $\VV_Q$, $Q\in \h^*$
as follows 
\beq
\VV_{\hat{\lambda}}\simeq
\bigoplus_{\mu\in\Gamma_k}\VV_{\hat\lambda,\mu}\otimes\VV^{\rm ext}_{\mu}
\simeq
\bigoplus_{\mu\in\Gamma_k}\VV_{\hat\lambda,\mu}\otimes
\bigoplus_{\delta\in{\rm Q}_l(\g)}\VV_{(\mu+k\delta)}\,, 
\eeq
where $\Gamma_k:={\rm P}(\g)/k{\rm Q}_l(\g)$ is a finite abelian group,
$\VV_{\hat\lambda,\mu}$ is a highest weight module 
of a generalized parafermionic $W$-algebra associated to the coset
construction $\hat{\g}_k/\hat{\h}$,   
and $\VV^{\rm ext}_\mu$ are highest weight modules with respect to an
extended $\hat{\h}$ $W$-algebra 
\cite{Gepner:1987sm}
\footnote{\label{stringfunction}In terms of characters
  $\chi_{\hat\lambda}(q,w)$  
corresponding to the $\hat{\g}_k$-highest weight representations, 
this is just the string function decomposition \cite{Kac:mq} 
$$
\chi_{\hat{\lambda}}(q,w)=\sum_{\mu\in\Gamma_k}
c_\mu^{\hat{\lambda}}(q) \theta_\mu(q,w)\,,\qquad 
\theta_\mu(q,w):=\sum_{\delta\in{\rm Q}_l(\g)}q^{{1\over 2}\kappa(\mu +
  k\delta,\mu+k\delta)} e^{2\pi i w(\mu+k\delta)}\,, 
$$
with $c_\mu^{\hat{\lambda}}$ the string functions and $\eta$ the
Dedekind-$\eta$-function.}. 

From this, the charge lattice can be read off to be 
\beq\el{chargelattice}
\Lambda_0=\{(\mu,\ol{\mu})\in{\rm P}(\g)\times{\rm P}(\g)\,|\,\mu-\ol{\mu}\in {\rm Q}(\g)\}\,.
\eeq
The ``duality group'' is given by the automorphisms
of $\Lambda_0$ compatible with the additional structures $S_k$, alluded to in the last section.
In the case of diagonal WZW models discussed here, all these structure
are determined by  
representation theory, and the duality group is given by the 
semi-direct product 
\beq
{\rm Aut}(\Lambda_0,S_k)\simeq {\rm A}(\g)\ltimes {\rm W}(\g)
\eeq
of the automorphism group ${\rm A}(\g)$ of the root lattice ${\rm
  Q}(\g)$ with the   
Weyl group ${\rm W}(\g)$, where ${\rm A}(\g)$ acts diagonally on
$\Lambda_0\subset{\rm P}(\g)\times{\rm P}(\g)$, 
and ${\rm W}(\g)$ acts on the second factor only (see
\cite{Kiritsis:1993ju} for a discussion 
of dualities of WZW models). 
Since 
${\rm A}(\g)\simeq {\rm W}(\g)\ltimes {\rm F}(\g)$, the ``duality''
groups can be written as  
\beq\el{gduality}
{\rm Aut}(\Lambda_0,S_k)\simeq \left({\rm W}(\g)\times{\rm
  W}(\g)\right)\ltimes {\rm F}(\g),  
\eeq
with the Weyl groups acting separately on the two factors of
$\Omega\times\Omega$ and ${\rm F}(\g)$ 
acting diagonally.
Note that these groups are finite, as opposed to the ``duality groups'' of
$d$-dimensional toroidal models for $d>1$. For the special case
$\g={\rm su}(2)$, the group 
\eq{gduality} coincide with the toroidal ``duality'' group ${\rm O}(1,1,\ZZ)$.

Given the duality group, the moduli space of current-current deformed
WZW models can now be written as
\eq{tormod}
\beq
{\cal M}_{\rm WZW}\simeq \left({\rm W}(\g)\times{\rm
  W}(\g)\right)\ltimes {\rm F}(\g)\backslash {\rm O}(d,d)/{\rm 
  O}(d)\times {\rm O}(d)\,. 
\eeq
In fact, $\Lambda_0$ has an even integral selfdual sublattice
\beq
\Lambda_k:=\{(\mu,\ol{\mu})\in{\rm P}(\g)\times{\rm P}(\g)\,|\,\mu-\ol{\mu}\in
k{\rm Q}_l(\g)\}\subset\Lambda_0
\eeq
of signature $(d,d)$, which can 
be regarded as charge lattice of $d$-dimensional
toroidal conformal field theory. 

Let us denote by ${\rm Aut}(\Lambda_k,S_k)\subset{\rm
  Aut}(\Lambda_0,S_k)$ 
the subgroup of the duality group fixing $\Lambda_k$. This group is
  also a subgroup of ${\rm O}(d,d,\ZZ)$. 

Every even integral selfdual signature
$(d,d)$ sublattice of $\Lambda_0$, which  
is obtained from $\Lambda_k$ by applying a transformation of ${\rm
  Aut}(\Lambda_0,S_k)$ 
gives rise to a representation of the WZW model as orbifold model
\beq\el{orbrep}
\hat{\g}_k\simeq\left(\hat{\g}_k/\hat{\h}\otimes {\rm
t_{\Lambda_k}}\right)\!\!/\Gamma_k\,. 
\eeq
of a product of a
coset model $\hat{\g}_k/\hat{\h}$ and a toroidal conformal field theory 
${\rm t}_{\Lambda_k}$ with charge lattices $\Lambda_k$.
Such representations were presented in \cite{Gepner:1987sm}
and were actually used in
  \cite{Kiritsis:1991zt} in the study of dualities of WZW and coset models.

The torus partition function of coset and toroidal models (with the
notation from footnote ${}^{\ref{stringfunction}}$)  
are given by
\beqn\el{pfpart} 
Z^{\hat{\g}_k/\hat{\h}}(q,\ol{q})=\sum_{\hat{\lambda}\in\hat{\Omega}_k}
\sum_{\mu\in\Gamma_k}    
|\eta(q)|^{2d} c^{\hat{\lambda}}_\mu(q) \ol{c}^{\hat{\lambda}}_\mu(\ol{q})\,,\\
\el{tpart}
Z^{{\rm t}_{\Lambda_k}}(q,\ol{q})=
\sum_{\mu\in\Gamma_k}{\theta_{\mu}(q) \ol{\theta}_{\mu}(\ol{q})
  \over|\eta(q)|^{2d}}   
=\sum_{(\mu,\ol{\mu})\in\Lambda} { q^{\mu^2\over 2k}
  \ol{q}^{\ol{\mu}^2\over 2k}\over |\eta(q)|^{2d}}\,. 
\eeqn
$\Gamma_k$ acts on the Hilbert spaces of the models by
\beqn\el{Gammaaction}
\gamma\in\Gamma_k:\quad 
\gamma\big|_{\VV_{\hat{\lambda},\mu}\otimes\ol{\VV}_{\hat{\lambda},\ol{\mu}}}
&=&e^{\pi i\kappa(\gamma,\mu+\ol{\mu})} {\rm
  id}_{\VV_{\hat{\lambda},\mu} \otimes\ol{\VV}_{\hat{\lambda},\ol{\mu}}}\,,\\
\gamma\big|_{\VV_{\mu}^{\rm ext}\otimes\ol{\VV}_{\ol{\mu}}^{\rm ext}}
&=&e^{-\pi i\kappa(\gamma,\mu+\ol{\mu})} {\rm id}_{\VV_{\mu}^{\rm ext}
  \otimes\ol{\VV}_{\ol{\mu}}^{\rm ext}}\,,\nonumber
\eeqn
giving rise to the $(\alpha,\beta)$-twisted torus partition functions
for $\alpha,\beta\in\Gamma_k$ 
\beqn\el{twcospart}
Z^{\hat{\g}_k/\hat{\h}}_{\alpha,\beta}(q,\ol{q})&=&
\sum_{\hat{\lambda}\in\hat{\Omega}_k}\sum_{\mu\in\Gamma_k} 
e^{\pi i\kappa(\alpha, 2\mu - \beta)}|\eta(q)|^{2d} c^{\hat{\lambda}}_\mu (q) 
\ol{c}^{\hat{\lambda}}_{\mu -\beta}(\ol{q})\,,\\
\el{twtpart}
Z^{{\rm t}_{\Lambda}}_{\alpha,\beta}(q,\ol{q})&=&
\sum_{\mu\in\Gamma_k} e^{-\pi i\kappa(\alpha, 2\mu-\beta)}
{\theta_\mu(q)\ol{\theta}_{\mu-\beta}(\ol{q})\over |\eta(q)|^{2d}}\,.
\eeqn
From this, one can easily read off that the orbifold partition function of
\eq{orbrep}
\beq
Z^{\rm orb}(q,\ol{q})={1\over |\Gamma_k|}\sum_{\alpha,\beta\in\Gamma_k}
Z^{\hat{\g}_k/\hat{\h}}_{\alpha,\beta}(q,\ol{q})Z^{{\rm
    t}_{\Lambda}}_{\alpha, \beta}(q,\ol{q})
\eeq
agrees with the torus partition function of the diagonal $\hat{\g}_k$-WZW model
\beq
Z^{\hat{\g}_k}(q,\ol{q})=\sum_{\hat{\lambda}\in\hat{\Omega}_k}
\chi_{\hat{\lambda}}(q)\chi_{\hat{\lambda}}(\ol{q})\,. 
\eeq
The fact that the orbifold group $\Gamma_k$ acts trivially on the
$W$-algebras of coset and toroidal models 
makes this representation of the WZW models useful for the study of
current-current deformations. 
Namely, for given $(\h,\ol{\h})$ the holomorphic and antiholomorphic
$W$-algebras  
$\hat{\h}\subset \hat{\g}_k$ and $\hat{\ol{\h}}\subset\hat{\g}_k$ 
can be identified with the ${\rm\hat{u}}(1)^d$, $\ol{{\rm\hat{u}}}(1)^d$ $W$-algebras of
${\rm t}_{\Lambda_k}$,  
and deformations corresponding to $(\h,\ol{\h})$ are then just
deformations of the toroidal factor ${\rm t}_{\Lambda_k}$ in \eq{orbrep}. 

Thus, for a point $O\in{\cal M}_{\rm WZW}$ the corresponding conformal
field theory can be represented as\footnote{A realization of the torus
partition function of the deformed ${\rm\hat{su}}(2)_k$-WZW models  
as a partition function of $\ZZ_k$-orbifolds of tensor product of
parafermionic models and 
free bosonic theories  
has already been given in \cite{Yang:1988bi}.}
\beq\el{deforb}
\hat{\g}_k(O)\simeq\left(\hat{\g}_k/\hat{\h}\otimes {\rm
  t_{O\Lambda_k}}\right)\!\!/\Gamma_k\,. 
\eeq
This provides concrete isomorphisms between 
current-current deformation spaces of WZW- and toroidal models.
Note however that in general the ``duality'' groups of the WZW models ${\rm\hat{g}}_k$ 
and the toroidal models ${\rm t}_{\Lambda_k}$ do not agree. On the one hand, there might
be dualities in the WZW models which do not preserve $\Lambda_k$ and thus correspond 
to a change of the orbifold representation, and on the other hand dualities in the 
toroidal model need not lift to automorphisms of $\Lambda_0$ together with $S_k$. As noted above,
for ${\g}={\rm su}(2)$, the duality group of the WZW model and the
corresponding toroidal models coincide. 

In fact, the conformal field theories in ${\cal M}_{\rm WZW}$ have WZW-like 
sigma model descriptions, which will be discussed in section \ref{sigmamodel}.
\section{Sigma Model Considerations}
\label{sigmamodel}
\subsection{Orbifold, Current-Current Deformation and
  Coset}\label{viereins} 
To show that the deformed WZW models discussed above indeed have
WZW-like sigma model descriptions, i.e.\ they are described by WZW-type
actions,  
however with metrics different from the bi-invariant metrics and
additional $B$-fields, 
we write down a sigma model action, which on the one hand defines
orbifold models as in \eq{deforb} and on the other hand describes
WZW-like models.

Let us start by describing the ingredients used to construct this sigma model.
As above, we denote by ${\rm G}$ a compact semi-simple Lie group of
rank $d$ with Lie algebra 
$\g$, Cartan subgroup ${\rm H}\subset {\rm G}$, corresponding Cartan
subalgebra $\h\subset\g$  
and $k\in\NN$. Furthermore,
$\Sigma$ is a $2$-dimensional surface, bounding the three manifold
$B$. Then, an  
{\it asymmetrically gauged WZW model} on $\Sigma$ is described by the
following action 
\beq\el{agWZW}
S^{\rm asym}_{{\rm G},k}(g,{\cal A},{\cal B},\lambda):= S_{{\rm G},k}^{\rm
  WZW}(g)+S_{{\rm G},k}^{\rm vg}(g,{\cal A})+S_{{\rm G},k}^{\rm
  ag}(g,{\cal B})  
+S_{{\rm G},k}^{\rm c}(g,{\cal A},{\cal B})+S_{{\rm G},k}^{\rm
  L}(\lambda,{\cal B})\,, 
\eeq
where
\beq\el{WZW}
S_{{\rm G},k}^{\rm WZW}(g)=-{k\over 4\pi}\left(\int_\Sigma d^2z \bra g^{-1}
\partial g, g^{-1}\ol{\partial}g\ket-{1\over 3}\int_B g^*\chi\right)
\eeq
is the WZW action with $g:\Sigma\longrightarrow {\rm G}$, $\bra
.,.\ket$ the Killing 
form on $\g$, and $\chi$ the three form on ${\rm G}$
associated to $\bra[.,.],.\ket$ \cite{Gepner:1986wi}. The integral
over $B$ in 
\eq{WZW} is called 
{\it Wess-Zumino term}. \eq{WZW} defines a quantum field theory 
if $H_2(G)=0$ and ${1\over 12\pi}\chi\in H^3(G,2\pi\ZZ)$,
which we assume to hold. 
This WZW model is 
vectorially coupled (see e.g.\ \cite{Gawedzki:1988nj}) to an ${\rm
  H}$-gauge connection ${\cal 
  A}=(A,\ol{A})\in\Omega^1(\Sigma,\h)$ by 
\beq\el{vectorpart}
S^{\rm vg}_{{\rm G},k}(g,{\cal A})=\frac{k}{2\pi} \int_\Sigma d^2 z\,
\left\{ \left< A, 
    g^{-1} \partial_{\bar{z}} g\right> - \left< \partial_z g g^{-1},
    \bar{A}\right> - \left< \left(1- \mbox{Ad}_g\right) A,
    \bar{A}\right> \right\} ,
\eeq
and axially coupled (see e.g.\ \cite{Kiritsis:1993ju,Ginsparg:1992af})
to an ${\rm 
  H}$-gauge connection ${\cal 
  B}=(B,\bar{B})\in\Omega^1(\Sigma,\h)$ by 
\beq\el{axialpart}
S^{\rm ag}_{{\rm G},k}(g,{\cal B})= \frac{k}{2\pi} \int_\Sigma d^2 z \left\{
\left< B, 
    g^{-1}\partial_{\bar{z}} g\right> + \left< \partial_z g g^{-1} ,
    \bar{B}\right> -    \left< \left( 1 +\mbox{Ad}_g\right) B,
    \bar{B}\right> \right\} \,.
\eeq
Adding both terms \eq{vectorpart} and \eq{axialpart} to the WZW
action \eq{WZW}, one obtains the general asymmetrically gauged model,
provided we introduce a coupling between the two gauge fields ${\cal
  A}$ and ${\cal B}$ 
\beq\el{coupling}
S^{\rm c}_{{\rm G},k}(g,{\cal A},{\cal B})=\frac{k}{2\pi} \int_\Sigma d^2 z
\left\{\left<\left( 1+ 
    \mbox{Ad}_g\right) B,
    \bar{A}\right>  -\left< \left( 1+\mbox{Ad}_g\right) A,
    \bar{B}\right> \right\} .
\eeq
This contains more terms than the interaction given
in \cite{Bars:1991pt}, where however a constraint relating the two gauge
fields was imposed. We avoid such a constraint by introducing a
Lagrange multiplier
\beq \el{lmultiplier}
S^{\rm L}_{{\rm G},k}\left( \lambda , {\cal B}\right) = \frac{ik}{\pi}
\int_\Sigma d^2z \, \left\{  
-\left< \partial_z \lambda ,\bar{B}\right> +\left< B,
    \partial_{\bar{z}}\lambda\right> \right\},
\eeq
with $\lambda:\Sigma\longrightarrow \h/(2\pi{\rm Q}_l(\g)^\prime)$, where
${\rm Q}_l(\g)^\prime$ is the 
image of ${\rm Q}_l(\g)$ under the isomorphism between $\h^*$ and $\h$
provided by $\bra .,.\ket$. 

This model will be coupled to an axially gauged ${\rm H}$-WZW model, i.e.
an axially gauged $d$-dimensional toroidal model
\beq \el{ypart}
S_{{\rm H},k,E}\left(y ,{\cal B}\right) =-\frac{k}{4\pi} \int_\Sigma d^2 z
\left< \left( 
    y^{-1}\partial_z y -2B\right), E\left( y^{-1}\partial_{\bar{z}}y
    -2\bar{B}\right) \right> ,
\eeq
where $y:\Sigma\rightarrow {\rm H}\simeq {\rm U}(1)^d$, and metric and $B$-field are
parametrized by an invertible $d\times d$-matrix with positive
definite symmetric part 
$E\in {\rm Gl}^{\rm ps}(d,\RR)$.

Below, it will be argued that the action
\beq\el{ausgangsmodell}
S_{{\rm G},k}(g,{\cal A},{\cal B},\lambda,y):=S_{{\rm G},k}^{\rm asym}(g,{\cal
  A},{\cal B},\lambda)+S_{{\rm H},k,E}(y,{\cal B}) 
\eeq
describes the current-current deformed $\hat{\g}_k$-WZW models discussed in 
section \ref{cftdefwzw}. In fact the vectorially gauged ${\rm G}$-WZW
model part 
$S^{\rm WZW}_{{\rm G},k}+S^{\rm vg}_{{\rm G},k}$ of the action
\eq{ausgangsmodell}  
describes the parafermionic factor
in the orbifold representation 
\eq{deforb} of the deformed models, while the ${\rm H}$-part
\eq{ypart} describes 
the toroidal one. The axial gauging which couples these two parts in
\eq{ausgangsmodell}  
amounts to the orbifold-construction in \eq{deforb}.

To see this, we will make use of the local symmetries of the model
\eq{ausgangsmodell} under vector transformations
\begin{eqnarray}
g & \to &  h g h^{-1}\\
A & \to & A + h\partial_z h^{-1} \nonumber\\
\bar{A} &\to& \bar{A} + h\partial_{\bar{z}} h^{-1} \nonumber\\
\lambda & \to & \lambda +\kappa ,\nonumber
\end{eqnarray}
with $h= \exp\left(i \kappa\right):\Sigma\longrightarrow {\rm H}$,
and axial transformations
\begin{eqnarray}
g & \to & fgf ,\\
B & \to & B + f^{-1}\partial_z f, \nonumber\\
\bar{B} & \to & \bar{B} + f^{-1}\partial_{\bar{z}} f, \nonumber\\
y &\to & fyf = f^2y =y f^2 ,\nonumber
\end{eqnarray}
with $f=\mbox{exp}(i\eta), \eta:\Sigma\longrightarrow \h/({1\over
  2}{\rm Q}(\g)^{\p})$. 

Let us first integrate out the Lagrange multiplier field $\lambda$.
Performing a partial integration in \eq{lmultiplier} yields
\beqn
S^{\rm L}_{{\rm G},k}(\lambda,{\cal B})
&=&\frac{ik}{\pi}\int_\Sigma\left(-d\bra \lambda, {\cal B}\ket
+\bra \lambda, {\cal F}({\cal B})\ket\right)\label{genorb}\\
&=& 2ik \sum_i \bra n_i,\oint_{\gamma_i}{\cal B}\ket
+{ik\over\pi}\int_\Sigma\bra \lambda, {\cal F}({\cal B})\ket\,.
\eeqn
where ${\cal F}=d{\cal B}$ is the curvature of ${\cal B}$, $\gamma_i$ represent
the non-trivial one-cycles of $\Sigma$ and $n_i$ are the ``winding
numbers'' of $\lambda$ 
around the dual cycles. Thus, the Lagrange multiplier forces ${\cal B}$ to be
flat with logarithms of monodromies around $\gamma_i$\footnote{A
  similar construction was used in \cite{Rocek:1991ps}.}  
\beq\el{monodromy}
b_i:=\oint_{\gamma_i}{\cal B}\in (2k{\rm Q}_l(\g))^*\simeq{1\over 2k}{\rm P}(\g)\,.
\eeq
Now, axial gauge transformations can shift the $b_i$ by elements in
${1\over 2}{\rm Q}_l(\g)$. 
Therefore, the gauge equivalence classes of flat connections ${\cal
  B}$ satisfying \eq{monodromy} 
are completely characterized by 
$b_i\in \left({1\over 2k}{\rm P}(\g)\right)/\left({1\over
  2}{\rm Q}_l(\g)\right)\simeq\Gamma_k$, 
and the integration over the gauge field ${\cal B}$ reduces to summing
over these $b_i$.  
Note that $\Gamma_k={\rm P}(\g)/(k{\rm Q}_l(\g))$ 
is nothing else than the orbifold group in the orbifold representation
of the deformed WZW models \eq{deforb}. 

Having integrated out $\lambda$ and ${\cal B}$ we end up with a
product of an ${\rm H}$-gauged ${\rm G}$-WZW model, corresponding to the
parafermionic coset model $\hat{\g}_k/\hat{\h}$ \cite{Gawedzki:1988nj},  
and a $d$-dimensional toroidal model parametrized by $E$, which are
coupled by $\Gamma_k$-twists. 
These twists indeed correspond to the $\Gamma_k$-action
\eq{Gammaaction} on the Hilbert spaces  
of the conformal field theories $\hat{g}_k/\hat{h}\otimes{\rm
  t}_{O_E\Lambda_k}$, and the 
sigma models \eq{ausgangsmodell} describe the deformed WZW models
\eq{deforb}. The identification of parametrizations of deformation spaces is 
as usual in toroidal CFTs: ${\rm O}(d,d)$ acts on ${\rm Gl}^{\rm ps}(d,\RR)$ by fractional
linear transformations, and we get $O({\bf 1})=E$ for $O\in{\rm O}(d,d)$ parametrizing
the orbifold models \eq{deforb} and $E\in {\rm Gl}^{ps}(d,\RR)$ parametrizing the sigma models
\eq{ausgangsmodell}. 

Next, we construct the ``T-dual'' models by integrating out ${\cal B}$
first. Since ${\cal B}$ enters the action algebraically, this can be
done by solving the equations of motion and plugging back the
solution into the action.
The calculation simplifies if we gauge fix $y=1$. Solving
the equations of motion for ${\cal B}$ yields
\beqn\el{bsol}
B &=& \left( \mbox{PAd}_g + 1 + 2E^T\right)^{-1} \left\{
  \mbox{P}\partial_z g 
  g^{-1} -\left( 1+ \mbox{Ad}_g\right) A - 2\Lambda^{-1}\partial_z
  \Lambda\right\}\,,\\
\bar{B} & = & \left( \mbox{PAd}_{g^{-1}} + 1 + 2E\right)^{-1} \left\{
  \mbox{P} g^{-1}\partial_{\bar{z}} g +\left(1
  +\mbox{Ad}_{g^{-1}}\right) \bar{A} + 2\Lambda^{-1}
  \partial_{\bar{z}}\Lambda\right\}\,,\nonumber
\eeqn
where ${\rm P}:\g\longrightarrow \h$ denotes the orthogonal projection on the
Cartan subalgebra  
and $\Lambda =\exp\left( i\lambda\right)$. This we plug back into the action
(\ref{ausgangsmodell}). In the end, we will also integrate over ${\cal A}$. 
Therefore, we gauge fix $\Lambda =1$ already at this stage. Then we obtain
\beqn\el{afterb}
S_{{\rm G},k}(g,{\cal A}) &=& S^{\rm WZW}_{{\rm G},k}(g) + S^{\rm
  vg}_{{\rm G},k}\left( g 
,{\cal A}\right)\\
&&
\qquad+{ \frac{k}{2\pi}\int_\Sigma d^2 
z\left( 
\left<  
    \mbox{PAd}_g + 1 + 
    2E^T\right)^{-1}\xi,\ol{\xi}\right>}\,,\nonumber\\
\xi &=& \mbox{P} 
    \partial_zg g^{-1} -\left( 
      1+\mbox{Ad}_g\right) A \,,\nonumber\\
\ol{\xi} &=& 
\mbox{P}
      g^{-1}\partial_{\bar{z}} g +\left( 1 +  
\mbox{Ad}_{g^{-1}}\right) 
  \bar{A}\,.\nonumber
\eeqn
Next, we integrate out ${\cal A}$ by solving the classical
equations of motion
\beqn
A & = & {\left( \mbox{PAd}_g -1 -\left( 1 +\mbox{PAd}_g\right) \left(
    \mbox{PAd}_g + 1 +2 E^T\right)^{-1} \left(1
    +\mbox{PAd}_g\right)\right)^{-1}}\nonumber\\
& & {\times \left\{ \mbox{P}\partial_zg g^{-1} -
    \left( 1 +\mbox{PAd}_g\right) \left( 
    \mbox{PAd}_g + 1 + 2E^T\right)^{-1} \mbox{P} \partial_z g
    g^{-1}\right\}}\,,\\         
\bar{A}& =& {\left( \mbox{PAd}_{g^{-1}} -1 -\left( 1
    +\mbox{PAd}_{g^{-1}}\right) \left( 
    \mbox{PAd}_{g^{-1}} + 1 + 2E\right)^{-1} \left(1
    +\mbox{PAd}_{g^{-1}}\right)\right)^{-1}}\nonumber\\
& & {\times\left\{- \mbox{P}g^{-1} \partial_{\bar{z}}g  +
    \left( 1 +\mbox{PAd}_{g^{-1}}\right) \left( 
    \mbox{PAd}_{g^{-1}} + 1 + 2E\right)^{-1} \mbox{P}g^{-1}
    \partial_{\bar{z}} g 
    \right\}}\,.\nonumber
\eeqn
Plugging this back into the action and performing some algebra yields
\beqn\el{wzwform}
S_{{\rm G},k}(g) &=& S^{\rm WZW}_{{\rm G},k} +\frac{k}{2\pi} \int d^2
z \left< \left( 
    \mbox{PAd}_g -R^{-1}\right)^{-1}
    \mbox{P} \partial_z g g^{-1} , \mbox{P} g^{-1} \partial_{\bar{z}} 
    g\right>\\
&=&-{k\over 4\pi}\int_\Sigma \bra {\cal
      E}_gg^{-1}\partial_zg,g^{-1}\partial_{\ol{z}}g\ket 
+ {k \over 12\pi} \int_B g^*\chi\,,\nonumber
\eeqn
with the abbreviations
\beqn
R &:=& {E^T-{\rm P}\over E^T+{\rm P}}\,,\\
{\cal E}_g &:=& (1-{\rm P})-\left({\rm P
  Ad}_g-R^{-1}\right)^{-1}\left({\rm P Ad}_g+R^{-1}\right)\\ 
&=& (1-{\rm P})+\left(R{\rm PAd}_g-{\rm P}\right)^{-1}
\left(R{\rm PAd}_g+{\rm P}\right)\,.\nonumber
\eeqn
Thus, we obtain the action of a WZW-like model with a deformed metric and
additional B-field encoded in the choice of $\h\subset\g$ and of the
matrix $E$. As expected from the comparison with the CFT considerations
we recover the action of the original ${\rm G}$-WZW model for $E={\bf 1}$.

Note that in general deformed metric and $B$-field are not
bi-invariant with respect to ${\rm G}$.  
But they are bi-invariant with respect to ${\rm H}\subset{\rm G}$,
which follows from the identities  
${\cal E}_{hg}={\cal E}_g$, ${\cal E}_{gh}={\rm Ad}_{h}^{-1}{\cal
  E}_g{\rm Ad}_h$ for all  
$g\in{\rm G}$, $h\in{\rm H}$.

Moreover, as also expected from the CFT results, 
for generic $E$ the model \eq{ausgangsmodell} only has a 
$\h\oplus\h$ chiral symmetry algebra which is generated by
\beq
\delta g = \epsilon g - g R\epsilon\,,\qquad
\bar{\delta} g = g\bar{\epsilon}-R^T{\bar{\epsilon}}g\,,
\eeq
with $\epsilon,\bar{\epsilon}:\Sigma\longrightarrow \h$.
The corresponding chiral currents read 
\beqn\el{hcurrents}
J & = & k R^{-1}\left( 1 -RR^T\right) \left(
\mbox{PAd}_g - R^{-1}\right)^{-1} \mbox{P} \partial_z g g^{-1} , \\
\bar{J} & = & - k  \left(R^{T}\right)^{-1}\left( 1- R^T R \right) \left(
\mbox{PAd}_{g^{-1}} - \left(R^{T}\right)^{-1}\right)^{-1} \mbox{P}
g^{-1}\partial_{\bar{z}} g \,.\nonumber
\eeqn
Let us mention that the two degenerations of the model \eq{ausgangsmodell}
$E=\lambda{\bf 1}$, $\lambda\rightarrow 0$, $\lambda\rightarrow\infty$
correspond to the axially and the vectorially gauged WZW model
respectively\footnote{This 
can be easily seen by comparing the sigma model
actions. Alternatively, it can be deduced by the following
observation. For $E=0$ the additional ${\rm U}(1)^d$ factor decouples and
integrating over ${\cal B}$ yields the ``T-dualized'' orbifold of the
vectorially gauged model, i.e.\ the axially gauged model. On the other
hand in the limit $\lambda\to \infty$ the gauge field ${\cal B}$ is frozen
to zero and we are left with the vectorially gauged model. In both
cases there is an additional decoupled ${\rm U}(1)^d$ factor whose torus is
of vanishing size or decompactified, respectively. In our previous
discussion this additional factor appears, because in the decoupling
limits the gauge fixing conditions need to be altered, i.e.\ in those 
limits one should gauge fix coordinates on ${\rm G}$ such that the metric on
the coset does not degenerate.}.
Thus, at the classical level, we achieved to construct a class of models
connecting the axially gauged WZW model via the ungauged to the
vectorially gauged one.

Moreover, the response
of our sigma model (\ref{wzwform}) to a variation of the deformation
parameters 
\begin{eqnarray}
\delta S & = & \frac{k}{2\pi} \int d^2 z \left<
\left(\delta R^{-1}\right) \left( \mbox{PAd}_g - R^{-1}\right)^{-1}
\mbox{P} \partial_z g g^{-1}, \left( \mbox{PAd}_{g^{-1}} -
\left(R^{T}\right)^{-1}\right)^{-1} \mbox{P}g^{-1}\partial_{\bar{z}}g \right>
\nonumber\\
& = & \frac{1}{2\pi k} \int d^2 z\left< R\left( R^TR-1\right)^{-1}
\left( \delta R^{-1}\right) \left( 1 - RR^T\right)^{-1} R J,
\bar{J}\right> 
\end{eqnarray}
is bilinear in the conserved currents \eq{hcurrents}, suggesting that
our family is indeed generated by current-current perturbations.

So far, we have seen that the action of the deformed model looks like
a WZW model action with deformed bilinear form, which is generically not bi-invariant
anymore. This however is not the full story. When integrating out the
gauge fields, one picks up Jacobians which usually give rise to a
non-trivial dilaton. Hence, we
expect that in addition to the deformed metric and $B$ field, there
will be a non-trivial dilaton. By construction our model should be
conformally invariant in the semiclassical limit. This means that the
background should satisfy beta function conditions (see e.g.\ the
review \cite{Tseytlin:1989md} and references therein). Checking these
equations one 
will also observe that a non trivial dilaton (coupling with the power
of $k^0$ to the sigma model action) is needed. That would be one way
to obtain the non-trivial dilaton. In the following subsection, we
will use a different method to derive the expression for the dilaton.

But before coming to that, let us comment on the relation of these 
deformed models to the families 
\beq\el{cosetaction}
S^{({\rm G}\times {\rm H})/{\rm H}}_{\tilde{E}}(g,{\cal
  B},y):=S_{{\rm G},k}^{WZW}(g)+S_{{\rm G},k}^{\rm ag}(g,{\cal
  B})+S_{{\rm H},k,\tilde{E}}(y,{\cal B}) 
\eeq
of gauged WZW models of type $({\rm G}\times {\rm H})/{\rm H}$ 
with varying embedding of the gauge group in the symmetry group of the
${\rm G}$- and ${\rm H}$-WZW models, parametrized by $\tilde{E}\in{\rm Gl}^{ps}(d,\RR)$. 
For the special case of symmetric $\tilde{E}$ these models were described 
in \cite{Tseytlin:1993hm}\footnote{
In fact, the concept of constructing families of 
sigma-models as gauged models with varying embedding of the gauge group
is very general in nature and has been applied e.g.\ to the case
of WZW-models corresponding to non-compact Lie groups in 
\cite{Horava:1991am,Nappi:1992kv}.}.

Integrating out the gauge fields ${\cal B}$ in these models,
one obtains\footnote{The result can be read off from \eq{bsol} 
by setting $A=\bar{A}=0$, $ E\mapsto \tilde{E}$ and
$\Lambda =1$.}
\beq\el{coset}
S^{({\rm G}\times {\rm H})/{\rm H}}_{\tilde{E}}(g)=S_{{\rm G},k}^{WZW}(g)+
 \frac{k}{2\pi} \int d^2z \left<
  \left( \mbox{PAd}_g +1 +2\tilde{E}^T\right)^{-1} \mbox{P} \partial_z
  g g^{-1} , \mbox{P} g^{-1}\partial_{\bar{z}} g\right>\,.
\eeq
This in fact coincides with the action \eq{wzwform}, which was
obtained by integrating out ${\cal B}$ in 
the sigma model \eq{ausgangsmodell} we started with, if one relates
the parameters 
\beq\el{relation-cos-orb}
\tilde{E} = - E\left( E -1\right)^{-1}\,.
\eeq
Hence, for $E$ such that $\tilde{E}$ in \eq{relation-cos-orb} is well-defined
and positive definite, the sigma model \eq{ausgangsmodell} we started with
has a coset realization given by \eq{cosetaction}. This is the case e.g.
for $E\in {\rm Gl}^{ps}(d,\RR)$ whose eigenvalues lie in $(0,1)$.
Note however that e.g.\ for $E=1$, $\tilde{E}$ in \eq{relation-cos-orb} is
not well-defined 
and thus, the original ${\rm G}$-WZW model does not have a proper
coset realization as in \eq{cosetaction}. 
It only corresponds to a degeneration thereof.
Nevertheless, for convenience, we will use the coset model realization
in the following subsection 
to calculate the Hamiltonian of the model
\eq{ausgangsmodell}. Although the realization we use 
is only defined on part of the actual moduli space, we suppose that
the result we obtain is  
actually valid on the whole moduli space. This is supported by the
observation that it reproduces the Hamiltonian of the WZW model at
$E={\bf 1}$. 
\subsection{Hamiltonian and Dilaton Shift}
Next, we would like to derive the Hamiltonian of the coset models
\eq{cosetaction}. 
In order to perform the Legendre transform, we
return to Minkowskian worldsheet signature, i.e.\ $z \to x^+ = \tau
+\sigma$, $\bar{z} \to x^{-} = \tau -\sigma$, $\partial_{z/
    \bar{z}} \to \partial_{+/-} = \frac{1}{2}\left( \partial_\tau \pm
  \partial_\sigma\right)$ and $d^2 z \to 2 d^2\sigma$. Let us first
discuss the ungauged ${\rm G}$-WZW model \eq{WZW}. The conjugate momenta are given by
\beq
\varpi^T = \left( \frac{\delta S}{\delta g^{-1} \partial_\tau
    g}\right)^T = -\frac{k}{4\pi} g^{-1}\partial_\tau g + \Omega\,, 
\eeq
where $\Omega$ denotes the contribution from the Wess--Zumino
term. Since it contains exactly one $\tau$ derivative,
its final effect drops out in the Hamiltonian. The Hamiltonian is
obtained by performing the Legendre transform
\begin{eqnarray}
H & = & \int d\sigma \mbox{Tr} \left( \varpi^T g^{-1}\partial_\tau
  g\right) -\int d\sigma L \nonumber\\
& = & -\frac{k}{8\pi} \int d\sigma \left( \left< g^{-1}\partial_\tau g,
  g^{-1}\partial_\tau g\right>  +  
\left< g^{-1}\partial_\sigma g, g^{-1}\partial_\sigma g\right>\right)
  ,
\end{eqnarray}
where $L$ stands for the Lagrangian belonging to \eq{WZW}. 
Introducing the currents
\beq
J_+ = -k \partial_+ g g^{-1} \,\,\, ,\,\,\, J_- = k g^{-1}\partial_- g 
\eeq
we arrive at the classical Sugawara expression
\beq\label{sugawara}
H = -\frac{1}{4\pi k} \int d\sigma \left( \left< J_+,J_+\right> + \left<
    J_-, J_-\right> \right) .
\eeq
For later use, we also give these currents expressed as phase space
functions
\begin{eqnarray}
J_+ & = & 2\pi \mbox{Ad}_g \varpi^T -2\pi k
\mbox{Ad}_g 
\Omega -\frac{k}{2} \partial_\sigma g g^{-1} ,\\
J_- &=& - 2\pi k \varpi^T +2\pi k\Omega -\frac{k}{2}
g^{-1}\partial_\sigma g .
\end{eqnarray}
In order to construct the Hamiltonian of the deformed ${\rm
  G}$-WZW models we follow 
the prescription given in \cite{Bowcock:xr}, where our starting point
  will be the 
coset model \eq{cosetaction} before integrating out the gauge fields,
  but after gauge 
fixing $y=1$, i.e. 
\begin{eqnarray}\el{axgamod}
S^{({\rm G}\times {\rm H})/{\rm H}}_{\tilde{E}}(g,{\cal B}) & = &
    S^{WZW}\!\left( g\right) +\frac{k}{\pi} \int\! 
    d^2 \sigma \bigg\{ \left< 
    B_+, g^{-1}\partial_{-}g\right> + \left< \partial_+ g
    g^{-1},B_-\right>\nonumber\\ 
    & &  \qquad\qquad\qquad\qquad\qquad -\left< \left( 1\! +\!
    \mbox{Ad}_g \! +\! 
    2\tilde{E}^T\right) B_+, B_-\right> \bigg\}\,.
\end{eqnarray}
The additional terms as compared to the ungauged model modify the
conjugate momenta according to
\beq
\varpi^T \longmapsto \varpi^T + \frac{k}{2\pi} B_+ + \frac{k}{2\pi}
\mbox{Ad}_g B_- .
\eeq
The corresponding Hamiltonian reads
\begin{eqnarray}
{\cal H} & = & H -\frac{k}{2\pi} \int d\sigma \bigg( -\left< B_+,
    g^{-1}\partial_\sigma g\right> +\left< \partial_\sigma g g^{-1} ,
    B_-\right>\nonumber\\ & & \qquad -2\left< \left( 1 + \mbox{Ad}_g + 2
    \tilde{E}^T\right) B_+ , 
    B_-\right> \bigg) ,
\end{eqnarray}
where $H$ denotes the Hamiltonian \eq{sugawara} of the ungauged model.

Next, we modify the currents $J_\pm$, such that
the modified currents
\begin{eqnarray}
{\cal J}_+ &=& J_+ + k\mbox{PAd}_g B_+ + kB_- ,\\
{\cal J}_- & = & J_- - kB_+ -k\mbox{PAd}_{g^{-1}} B_- .
\end{eqnarray}
obey a gauge invariant Poisson algebra (see \cite{Bowcock:xr} for more details).
In terms of the new currents the Hamiltonian of the deformed model is given by
\begin{eqnarray}\el{hwithb}
{\cal H} & = & -\frac{1}{4k\pi} \int d\sigma \Big( \left< {\cal
      J}_+,{\cal J}_+\right> + \left< {\cal J}_- , {\cal J}_-\right>
-  4k\left< B_-, {\cal
      J}_+\right> + 4k\left< B_+, {\cal J}_-\right>
      \nonumber\\ 
& &  \,\,\,\,\,\, +2k^2\left< B_+ ,
      B_+\right> +2k^2\left< B_- ,B_-\right> - 4k^2\left<B_+, \left(
      1+2\tilde{E}\right) B_-\right>\Big)\,.
\end{eqnarray}
The additional constraints due to the vanishing of the conjugate
momenta of $B_{\pm}$ 
can be used to eliminate the gauge fields by solving
their algebraic equations of motion
\begin{eqnarray}
\left( 1 + 2\tilde{E}^T\right) B_+ - B_- = -\frac{1}{k}\mbox{P}{\cal
  J}_+ ,\label{algbone}\\ 
\left( 1+2\tilde{E}\right) B_- - B_+ = \frac{1}{k}\mbox{P}{\cal J}_-
  .\label{algbtwo} 
\end{eqnarray}
For $\tilde{E}$ invertible\footnote{The degeneration $\tilde{E}=0$ for example
describes the
  coset model ${\rm G}/{\rm H}$ which 
  is discussed in \cite{Bowcock:xr}.} these equations allow
  a unique 
  solution for $B_\pm$. Before giving this it is useful to employ
  (\ref{algbone}) and (\ref{algbtwo}) to simplify the expression
  (\ref{hwithb}) slightly. For the last term we write
$$ 2\left< \left( 1+2\tilde{E}^T\right) B_+, B_-\right> = \left<\left(
    -\frac{1}{k}\mbox{P}{\cal J}_+ + B_-\right), B_-\right> +\left< B_+,
  \left(\frac{1}{k}\mbox{P}{\cal J}_- + B_+\right)\right> $$
and obtain
\beq
{\cal H} = -\frac{1}{4\pi k}\int d\sigma \Big( \left< {\cal J}_+,{\cal
    J}_+\right> + \left< {\cal J}_-, {\cal J}_-\right> +2k\left< B_+,
    {\cal J}_-  \right> -2k\left< B_-, {\cal J}_+\right>\Big) .
\eeq
Finally, we plug in the solutions of the equations of motion
(\ref{algbone}), (\ref{algbtwo}) of ${\cal B}$
\begin{eqnarray}
B_+ & = & \frac{1}{2k}\left( \tilde{E} + \tilde{E}^T +
  2\tilde{E}\tilde{E}^T\right)^{-1} \left(\mbox{P} 
  {\cal J}_- -\left( 1 + 2\tilde{E}\right) \mbox{P}{\cal J}_+\right) ,\\
B_- & = & \frac{1}{2k} \left( \tilde{E} + \tilde{E}^T + 2\tilde{E}^T
  \tilde{E}\right)^{-1} \left( 
  \left( 1 + 2\tilde{E}^T\right) \mbox{P}{\cal J}_- -2\mbox{P}{\cal J}_+\right)
  .
\end{eqnarray}
The result for the Hamiltonian of the deformed model is
\begin{eqnarray}
{\cal H} & = & -\frac{1}{4\pi k} \int d\sigma \Bigg( \left< \Big( 1
    +\mbox{P}\left( 
    \tilde{E} + \tilde{E}^T +2\tilde{E}^T\tilde{E}\right)^{-1}
    \mbox{P}\Big) {\cal J}_+ , {\cal 
    J}_+\right>\nonumber\\ & & \quad 
+\left< \left( 1 +\mbox{P}\left(
    \tilde{E} + \tilde{E}^T +2\tilde{E}\tilde{E}^T\right)^{-1}
    \mbox{P}\right) {\cal J}_- , {\cal 
    J}_-\right> \nonumber\\
& & -\left< \Big( \left( \tilde{E} + \tilde{E}^T\!
    +2\tilde{E}\tilde{E}^T\right)^{-1}\left(1+2\tilde{E}\right)
\right. \nonumber\\ & &  \left.   
    +\left( 1+2\tilde{E}\right)\left( \tilde{E} + \tilde{E}^T\! +
    2\tilde{E}^T\tilde{E}\right)^{-1}\Big) 
    \mbox{P} {\cal
    J}_+, \mbox{P}{\cal J}_-\right> \Bigg). \label{finalH} 
\end{eqnarray}
For symmetric $\tilde{E}$ this expression agrees with the one given
in \cite{Tseytlin:1993hm}.

Now, the zero mode part of the Hamiltonian restricted to a certain subspace of the Hilbert space 
should ``match'' with the generalized Laplacian on the space of smooth functions on ${\rm G}$ \cite{Tseytlin:1993my}
\beq\el{genlaplace}
\Delta^\Phi:=-\frac{e^{2\Phi}}{\sqrt{{\rm det}(G)}}\partial_\mu
  e^{-2\Phi}\sqrt{{\rm det}(G)}G^{\mu\nu} \partial_\nu\,,
\eeq
which takes into account the dilaton $\Phi$.
This can be explained (for more details see \cite{Tseytlin:1993my}) 
by the correspondence of the 
constraint 
\beq\label{constr}
\left( L_0 +\bar{L}_0 - a\right)\left| \mbox{physical}\right> = 0 
\eeq
($L_0 +\bar{L}_0$ is the Hamiltonian and
$a$ is a normal ordering constant)
with the mass shell condition, which in location space reads
\beq\label{massshell}
\frac{e^{2\Phi}}{\sqrt{{\rm det}(G)}}\partial_\mu \left(
  e^{-2\Phi} 
  \sqrt{{\rm det}(G)}G^{\mu\nu} \partial_\nu\Psi\right) = -m^2 \Psi \,.
\eeq
Here $\Psi$ denotes the target space field associated to the physical
state in (\ref{constr}). This will be used in the following to determine the
dilaton in the deformed ${\rm G}$-WZW models.

The subspace of the Hilbert space which should correspond to the space of functions
on ${\rm G}$ consists of highest weight vectors only (see \cite{Frohlich:1993es}
for a discussion of this point), and thus the Hamiltonian \eq{finalH} 
restricted to this subspace can be written completely in terms of zero-modes 
of left and right currents. 
Since zero-modes of left and right chiral currents are the generators
of left- and inverse right- 
multiplication by ${\rm G}$ on itself, they can be identified with the
respective sections 
$j_L$ and $j_R$ of ${\rm TG}\otimes\g^*$. Choosing a basis of $\g^*$,
one obtains from these 
sections the vector fields $j^A_L$, $j^A_R$, $A\in\{1,\ldots,{\rm
  dim}({\rm G})\}$ on ${\rm G}$. 
In every point $p\in{\rm G}$, $((j_L^A)_p)_{A}$, $((j_R^A)_p)_{A}$ are
two basis  
of the tangent space ${\rm T}_p{\rm G}$ of ${\rm G}$ in $p$. Hence,
$\Delta^\Phi$ can be written 
in terms of $j_L^A$ or $j_R^A$, $\Phi$ and the target space metric $G$. 
$G$ can be read off from the kinetic term
of the action \eq{wzwform}, such that we can compare $\Delta^\Phi$
with the Hamilton operator 
\eq{finalH} to obtain $\Phi$. This is the general strategy presented
in \cite{Tseytlin:1993my}.  

But before applying it to the models \eq{ausgangsmodell}, let us illustrate it 
in the example of the ``undeformed'' ${\rm G}$-WZW model.
For notational convenience, we will use the sections 
$j_L$ and $j_R$ of ${\rm TG}\otimes\g^*$
in the following computations and denote the dual Killing
form on $\g^*$ by $\bra.,.\ket$.
Now, the target space metric is $2\pi k$ times the bi-invariant metric
induced by the Killing form. The Laplace 
operator can then be written as
\beq
\Delta= \frac{1}{k} \bra j_L , j_L \ket =
\frac{1}{k} \bra j_R ,j_R \ket =
\frac{1}{2k}\left( \bra j_L, j_L \ket + \bra j_R, j_R \ket\right)\,,
\eeq 
where the second equality
reflects the bi-invariance of the metric.

Observing that under the identification $J^A_0\sim j_L^A$,
$\bar{J}^A_0\sim j_R^A$, 
of zero-modes of the holomorphic and antiholomorphic currents with
generators of the left and right 
multiplication the Hamiltonian \eq{sugawara} ``matches'' the 
Laplacian $\Delta$, we deduce that the dilaton $\Phi$ is constant.

Now let us come to the discussion of the deformed models
\eq{ausgangsmodell}. The Hamiltonian 
corresponding to the coset representation of these models has been
calculated above \eq{finalH}. 
The target space metric can be
  obtained from (\ref{coset}) by symmetrization\footnote{As noted
    before the actions 
\eq{ausgangsmodell} and \eq{coset} agree under the identification
\eq{relation-cos-orb}, 
and we will use the parametrization by $\tilde{E}$ for the calculation
of $\Delta^\Phi$,  
because the Hamiltonian \eq{finalH} was obtained in the coset
representation. This however 
does not restrict the region of validity of the results.}
\begin{eqnarray}
G_{\mu\nu}/{2\pi k} & = &\left< M j_{L\mu}, j_{L\nu}\right>
  \nonumber\\ 
& = & \left<N j_{R\mu}, j_{R\nu}\right>,
\end{eqnarray}
with
\begin{eqnarray}
M &  = & \left( 1\! +\! 2 \tilde{E}^T\mbox{P}\!
  +\!\mbox{PAd}_g\mbox{P}\right)^{-1} \left( 1\!-\!\mbox{P}\! +\!
  2\tilde{E}^T\mbox{P}\! + \!
  2 \tilde{E}\mbox{P} \! +\! 4 \tilde{E}^T \tilde{E}\mbox{P}
  \right)\nonumber\\ & & \quad\times\left(1\! 
  +\! 2 \tilde{E}\mbox{P} 
  \! +\!\mbox{PAd}_{g^{-1}}\mbox{P}\right)^{-1} , \\
\hspace*{-.3in}
N &  = & \left( 1\! +\! 2 \tilde{E}\mbox{P}\!
  +\!\mbox{PAd}_{g^{-1}}\mbox{P}\right)^{-1} \left( 1\!-\!\mbox{P}\! +\!
  2\tilde{E}^T\mbox{P}\! + \!
  2 \tilde{E}\mbox{P} \! +\! 4 \tilde{E} \tilde{E}^T\mbox{P}
  \right)\nonumber\\ & & \quad\times\left(1\! 
  +\! 2 \tilde{E}^T\mbox{P} 
  \! +\!\mbox{PAd}_{g}\mbox{P}\right)^{-1} .
\end{eqnarray}
Now, the generalized Laplacian can be written as
\beq
4\pi k \Delta^\Phi = \left< f^{-1} j_L , f M^{-1} j_L \right> + \left<
  f^{-1} j_R, fN^{-1}j_R \right> ,
\eeq
where derivatives act on everything appearing to their right, and
\beq
f= e^{-2\Phi} \sqrt{{\rm det}(G)}/\sqrt{{\rm det}(G_0)} ,
\eeq
with $G_0$ denoting the ``undeformed'', i.e.\ the Killing metric on
${\rm G}$.    
The expression for the inverse of $M$ is
\begin{eqnarray}
2M^{-1} &=& 2+ \mbox{P}\left( \tilde{E}^T + \tilde{E} + 2\tilde{E}
\tilde{E}^T\right)^{-1}\mbox{P} 
+\mbox{P}\left(1 + 2\tilde{E}\right)\left( \tilde{E}^T + \tilde{E} +
2\tilde{E}^T 
  \tilde{E}\right)^{-1}\mbox{PAd}_g\mbox{P} \nonumber\\ & & +  
\mbox{PAd}_{g^{-1}}\mbox{P}\left( \tilde{E}^T + \tilde{E} +
2\tilde{E}^T \tilde{E}\right)^{-1} \left(1 
    + 2\tilde{E}^T\right)\mbox{P} \nonumber\\ & & +
    \mbox{PAd}_{g^{-1}}\mbox{P} 
    \left( \tilde{E}^T + \tilde{E} + 2\tilde{E}^T
    \tilde{E}\right)^{-1} \mbox{PAd}_g\mbox{P} . 
\end{eqnarray}
In order to obtain this result we have performed a couple of
manipulations. First, we employed that 
$$ 
\left( 1\! +\! 2 \tilde{E}^T\mbox{P}\!
  +\!\mbox{PAd}_g\mbox{P}\right)^{-1}\left( 1\!-\!\mbox{P}\right)  =
\left( 1\!-\!\mbox{P}\right) 
\left(1\! +\! 2 \tilde{E}\mbox{P} 
  \! +\!\mbox{PAd}_{g^{-1}}\mbox{P}\right)^{-1} = 1-\mbox{P},
$$
which follows from P being a projector. Then $M$ takes a block diagonal form, and we
have inverted the two blocks ($1-P$ and the rest) on the corresponding
subspaces. 
Furthermore, on the Cartan subalgebra $\h$ (where P=1) we used
the following identities
\begin{eqnarray}
\lefteqn{-1 + \frac{1}{2} \left( 1\! +\! 2\tilde{E}\right)\left(
 \tilde{E}^T\! +\! \tilde{E}\! + 
 2\tilde{E}^T\tilde{E}\right)^{-1} \left( 1\!+\! 2\tilde{E}^T\right)
 =}& & \nonumber\\ & & 
\hspace*{-.3in} \frac{1}{2} 
 \left( 1\! +\! 2\tilde{E}\right)\left( \tilde{E}^T\! +\! \tilde{E}\! +\! 
 2\tilde{E}^T\tilde{E}\right)^{-1} \bigg( -2\left(\tilde{E}^T\! +\!
 \tilde{E}\! +\! 2\tilde{E}^T\tilde{E}\right)\! +\! 
 \left(1\! +\! 2\tilde{E}^T\right) 
 \left(1\! +\! 2\tilde{E}\right) \bigg) \left(1\! +\!
 2\tilde{E}\right)^{-1}\nonumber\\
& & = \frac{1}{2} \left( 1 +2\tilde{E}\right) \left( \tilde{E}^T +
 \tilde{E} + 2\tilde{E}^T 
 \tilde{E}\right)^{-1}\left( 1+2\tilde{E}\right)^{-1} \nonumber\\
& & = \frac{1}{2}\left( \tilde{E}^T + \tilde{E} + 2\tilde{E}
 \tilde{E}^T\right)^{-1} \label{manipu}\,.
\end{eqnarray}
The last line of (\ref{manipu}) is most easily checked with the
inverted expression
\begin{eqnarray}
\left( 1+ 2\tilde{E}\right) \left( \tilde{E}^T + \tilde{E} + 2\tilde{E}^T
 \tilde{E}\right) \left( 1+ 2\tilde{E}\right)^{-1}  & = & \left( 1+
 2\tilde{E}\right)\tilde{E}^T + 
 \tilde{E}\nonumber\\ & =& \tilde{E}^T + \tilde{E} + 2\tilde{E}\tilde{E}^T.
\end{eqnarray}
Analogously, one finds
\begin{eqnarray}
2N^{-1} &=& 2+ \mbox{P}\left( \tilde{E}^T + \tilde{E} + 2\tilde{E}^T
\tilde{E}\right)^{-1}\mbox{P} 
+\mbox{P}\left(1 + 2\tilde{E}^T\right)\left( \tilde{E}^T + \tilde{E} +
2\tilde{E} 
  \tilde{E}^T\right)^{-1}\mbox{PAd}_{g^{-1}}\mbox{P} \nonumber\\ & &
  \hspace*{-1in}+   
\mbox{PAd}_{g}\mbox{P}\left( \tilde{E}^T + \tilde{E} + 2\tilde{E}
\tilde{E}^T\right)^{-1} \left(1 
    + 2\tilde{E}\right)\mbox{P} + \mbox{PAd}_{g}\mbox{P} \left(
    \tilde{E}^T + \tilde{E} + 
      2\tilde{E} \tilde{E}^T\right)^{-1} \mbox{PAd}_{g^{-1}}\mbox{P} .
\end{eqnarray}
In order to write down the result for the Laplacian, we also
need the relations
$$
\mbox{Ad}_g j_L = -j_R \,\,\, ,\,\,\, 
\mbox{Ad}_{g^{-1}} j_R = - j_L .
$$
Collecting everything, we finally obtain
\begin{eqnarray}
4 \pi k \Delta^\Phi & = & \left< f^{-1} j_L, f\left(
1+\mbox{P}\left( \tilde{E}^T\! 
+\! \tilde{E} \! +\!  2\tilde{E}\tilde{E}^T\right)^{-1}\mbox{P}\right)
j_L\right>\nonumber\\ 
& & +  \left< f^{-1}
j_R, f\left(  
1+\mbox{P}\left( \tilde{E}^T \! +\!  \tilde{E} + 2\tilde{E}^T\!
\tilde{E}\right)^{-1}\mbox{P}\right) 
j_R\right> 
\nonumber\\ 
& & - \frac{1}{2} \left< f^{-1} \mbox{P} j_L,f \left(
\left( 1\! + \!  
2\tilde{E}\right) \left( \tilde{E}^T\! +\! \tilde{E}\! +\! 2\tilde{E}^T \!
\tilde{E}\right)^{-1}\right.\right.\nonumber\\ & &
\hspace*{1in}\left.\left. + \left( 
\tilde{E}^T \!  +\! \tilde{E}\! +\!
2\tilde{E} \tilde{E}^T\right)^{-1} \left( 1\! +\!
2\tilde{E}\right)\right)\mbox{P} 
j_R\right> \nonumber\\
& & - \frac{1}{2} \left< f^{-1} \mbox{P} j_R,f \left(
\left( 1 \! + 
\! 2\tilde{E}^T\right) \left( \tilde{E}^T\! +\! \tilde{E}\! +\!
2\tilde{E} \tilde{E}^T\right)^{-1} 
\right.\right.\nonumber\\ & & 
\hspace*{1in}\left.\left. + \left(
\tilde{E}^T\! +\! \tilde{E}\! +\! 
2\tilde{E}^T\! \tilde{E}\right)^{-1} \left( 1\! +\!
2\tilde{E}^T\right)\right)\mbox{P} 
j_L\right>\! . \label{laplaceF}
\end{eqnarray}
This expression matches with the Hamiltonian (\ref{finalH}) provided
that $f$ is equal to a constant, which can be taken to one by
performing a constant dilaton shift. Thus, we conclude that along the
deformations there is a non trivial dilaton such that
\beq
\sqrt{{\rm det}(G)}e^{-2\Phi} \,\,\, \mbox{is independent of $\tilde{E}$} .
\eeq
The independence of this `string measure' on the deformation
parameter(s) has been observed before for one dimensional deformations
\cite{Hassan:1992gi,Giveon:1993ph,Sfetsos:1993ka} 
and for symmetric $\tilde{E}$ \cite{Tseytlin:1993hm}.

As a byproduct we see from \eq{laplaceF} that eigenfunctions of
$\Delta^\Phi$ at a given $\tilde{E}$,  
which are also eigenfunctions with respect to the ${\rm H}$-actions induced
by the left and right multiplication of ${\rm H}$ on ${\rm G}$ are
also eigenfunctions of $\Delta^\Phi$ 
at all $\tilde{E}$, with different 
eigenvalues however. This is expected from the CFT considerations above, 
where the only effect of the deformations were changes of the $\h$,
$\ol{\h}$ charges, and $\hat{\h}$-, $\hat{\ol{\h}}$-highest weight
states remained  
highest weight states under the deformations. 

To complete the discussion we should recall that the coset description
we used here to construct the Hamiltonian \eq{finalH} is only valid in the part of the 
parameter space of the model \eq{ausgangsmodell}, where $\tilde{E}$ is positive
definite. In particular, it
breaks down, when one of the eigenvalues of $E$ becomes $1$.
Nevertheless, the Hamiltonian, we obtained, can be continued to
the region where $E$ has eigenvalues $1$, as can be read off from
\begin{eqnarray}
{\cal H} & = & -\frac{1}{4\pi k} \int d\sigma \bigg( \left< \Big( 1
    +\mbox{P}\left( E-1\right)\left(
    E + E^T \right)^{-1}\left( E^T -1\right) 
    \mbox{P}\Big) {\cal J}_+ , {\cal 
    J}_+\right>\nonumber\\ & & \quad +
 \left< \left( 1 +\mbox{P}\left( E^T-1\right)\left(
    E + E^T \right)^{-1}\left( E -1\right)
    \mbox{P}\right) {\cal J}_- , {\cal 
    J}_-\right> \nonumber\\
& & +2\left< \left( E^T-1\right) \left( E+E^T \right)^{-1} \left(
    E+1\right)  
    \mbox{P} {\cal
    J}_+, \mbox{P}{\cal J}_-\right>\,, \label{finalHwE}
\end{eqnarray} 
and in fact for $E={\bf 1}$ it coincides with the one of the ${\rm G}$-WZW model 
\eq{sugawara}.
This suggests that \eq{finalHwE} is indeed the Hamiltonian on the whole parameter
space of \eq{ausgangsmodell} and all the results from this section also apply to
the entire moduli space.

\subsection{The ${\rm SU}(2)$ Example}

In this subsection we would like to illustrate our previous discussion very explicitly in 
the simplest example, namely ${\rm G}={\rm SU}(2)$. We do not want to go through
all the details, since much of the discussion for ${\rm SU}(2)$ (or
$SL(2,{\mathbb R})$)\footnote{The discussion of deformation is often
  presented for the non compact version of $A_1$ because in that case
  the interesting phenomenon of smooth topology change is
  observed.}  can be found in the literature
(e.g.\ in \cite{Hassan:1992gi,Giveon:1993ph,Sfetsos:1993ka}), but we will   
address two topics here, which were important in the previous sections. 
Firstly, we will derive the deformed sigma model 
by taking the T-dual of the model
$\left( {\rm SU}(2)_k/{\rm U}(1) \!\times\! {\rm U}(1)\right)/{\mathbb  Z}_k$, 
and secondly we will compute the spectrum of the generalized Laplacian.

From the discussion of current-current deformed WZW models corresponding
to arbitrary compact semi-simple Lie groups in section \ref{cftdefwzw}, 
we know that the deformed ${\rm\hat{su}}(2)_k$-WZW models can be realized as orbifold models
(compare \eq{deforb})
\beq
\left({\rm\hat{su}}(2)_k/{\rm\hat{u}}(1) \otimes
     {\rm\hat{u}}(1)_{\sqrt{k}R}\right)/{\mathbb  Z}_k\,, 
\eeq
where $R\in (0,\infty)$ parametrizes the
${\rm\hat{u}}(1)$-factor\footnote{As alluded to above, $R\mapsto {1\over R}$ is a
  duality.}.  
That the deformed ${\rm\hat{su}}(2)_k$-WZW model can be written in this way
has first been suggested  
by Yang in \cite{Yang:1988bi}, where 
a one-parameter family of modular invariant partition functions corresponding
to such orbifold models with varying radius in the ${\rm\hat{u}}(1)$-factor
was presented, 
noting that the partition function at $R=1$ coincides with the
partition function of the ${\rm\hat{su}}(2)_k$-WZW model
\cite{Gepner:1986hr}.  

In the following we will compare this with explicit sigma model
analysis. 
  We start with the $({\rm SU}(2)/{\rm U}(1) \times {\rm U}(1))/{\mathbb Z}_k$ gauged
  sigma model (the  
  parametrization is taken from \cite{Giveon:1993ph})\footnote{In
    order to make contact with our general discussion in
    subsection\ \ref{viereins} we note that in \cite{Giveon:1993ph} an 
    ${\rm SU}(2)$ group element is written as $\mbox{exp} \left[ i\left(\theta
      -\tilde{\theta}\right)\sigma_3 /2\right] \mbox{exp} \left[ ix
      \sigma_2\right] 
      \mbox{exp}\left[ i \left(\theta +\tilde{\theta}\right)
	\sigma_3/2\right]$. The dependence on $\tilde{\theta}$ is
      removed after performing the vector gauging. Constant shifts in
      $\theta$ correspond to the axial symmetry. The orbifold action
      of ${\mathbb Z}_k$ will be specified below in the discussion
      after (\ref{su2orb}). It essentially reduces the radius of a
      circle lying diagonally in the $\theta$$y$-torus.} 
\beq
S=\frac{k}{2\pi} \int d^2z\left\{ \partial_+ x\partial_- x + \tan^2x
  \partial_+ \theta \partial_- \theta + \frac{1}{R^2} \partial_+
  y\partial_- y\right\} ,
\eeq
where for the time being we omitted the dilaton term coupling to the
Gauss-Bonnet density. Now, we redefine coordinates according to
\beq
\theta =  \alpha + \beta \,\,\, ,\,\,\,
y =\alpha -\beta
\eeq
In these coordinates the relevant components of the target space metric can be written as
\begin{eqnarray}
G_{\alpha\alpha} & = & G_{\beta\beta} =k\left(  \tan^2x +
  \frac{1}{R^2}\right) \\
G_{\alpha\beta} & = & k\left( \tan^2x -
  \frac{1}{R^2}\right) . 
\end{eqnarray}
Now we T-dualize the $\alpha$ direction. 
The T-dual metric and $B$-field follow from the
Buscher fromul\ae\ \cite{Buscher:sk}. We obtain
\begin{eqnarray}
\tilde{G}_{\tilde{\alpha}\tilde{\alpha}} & =
    &\frac{1}{G_{\alpha\alpha}}= \frac{R^2\cos^2 x}{k\left( 
    \cos^2x + 
    R^2 \sin^2 x\right)} \\
\tilde{G}_{\beta\beta} & = & G_{\beta\beta} -
    \frac{G_{\alpha\beta}^2}{G_{\alpha\alpha}} = \frac{k\sin^2
    x}{\cos^2 x + R^2 \sin^2 x} \\
\tilde{G}_{\tilde{\alpha}\beta} &=&
    \frac{B_{\alpha\beta}}{G_{\alpha\alpha}} = 0\\
\tilde{B}_{\tilde{\alpha}\beta} &=&
    \frac{G_{\alpha\beta}}{G_{\alpha\alpha}} = -2\frac{\cos^2x
    }{\cos^2 x+ R^2\sin^2 x} +1 
\end{eqnarray}
Now we define $\tilde{\alpha}  = k\tilde{\theta}$, $\beta =\theta$ and
drop the constant term 1 in the $B$-field. 
This amounts to 
\begin{eqnarray}
S^R & = & \frac{k}{2\pi} \int d^2z \left\{ \partial_+ x\partial_- x +
  \frac{\sin^2 x}{\cos^2 x + R^2 \sin^2 x} \partial_+ \theta
  \partial_- \theta \right. \nonumber \\
& & \hspace{-1in}
\left. \; \; + \frac{ R^2 \cos^2 x}{\cos^2 x + R^2 \sin^2 x}
  \partial_+  
  \tilde{\theta} \partial_- \tilde{\theta}   +\frac{\cos^2 x}{\cos^2 x
  + R^2 \sin^2 x} \left( 
  \partial_+ 
  \theta \partial_- \tilde{\theta} - \partial_+
  \tilde{\theta}\partial_- \theta \right)\right\} .
\end{eqnarray}
It remains to discuss the periodicity of $\tilde{\theta}$. 
In order to obtain the deformed model $\tilde{\theta}$ should be a
$2\pi$ periodic coordinate, i.e.\ $\tilde{\alpha}$ should be a $2\pi
k$ periodic coordinate. If we perform the T-duality according to the
prescription given in \cite{Rocek:1991ps}, the intermediate gauged sigma
model on a worldsheet $\Sigma$ will contain a term (compare \eq{genorb})
\beq \label{su2orb}
S_{wind} = \sum_i n_i \oint_{\gamma_i} A \, , 
\eeq
where $\gamma_i$ are one cycles of $\Sigma$ and 
$n_i$ are the winding numbers of the Lagrange
multiplier $\tilde{\alpha}$ around one cycles of 
$\Sigma$ dual to $\gamma_i$. 
As in the discussion around \eq{genorb}, 
summing over the $k {\mathbb Z}$ valued windings of
$\tilde{\alpha}$ yields a Kronecker delta, which is non-vanishing if $\oint A$
takes values in $2\pi{\mathbb Z}/k$. Since the original model is
obtained by absorbing a pure gauge $A_{\pm} = \partial_{\pm} \rho$
into a redefinition of the
original coordinate, $\alpha$ parameterises actually an orbifold
$S^1/{\mathbb Z}_k$ where $S^1$ is the unit circle. 

Thus for the case ${\rm G}={\rm SU}(2)$ we explicitly obtained the 
deformed WZW action \eq{wzwform} from the orbifold representation
$\left({\rm SU}(2)_k/{\rm U}(1)\times {\rm U}(1)\right)/{\mathbb Z}_k$.

Next, we would like to discuss the generalized Laplacian $\Delta^\Phi$ in this
example. For notational simplicity we set $k=1$ during the
calculation and reinstall it in the end. 
After a coordinate change
\beq
\rho = \sin x
\eeq
the metric of the deformed model takes the form
\beq
ds^2 = \frac{d\rho^2}{1-\rho^2} +\frac{\rho^2}{1+\left( R^2 -1\right)\rho^2}
d\theta^2 + \frac{R^2\left( 1-\rho^2\right)}{1+\left( R^2 -1\right)
  \rho^2} d\tilde{\theta}^2 .
\eeq
Up to a constant shift, the dilaton is given by the relation
\beq
e^{-2\Phi}\sqrt{{\rm det}(G)} =\rho .
\eeq
Thus, we find explicitly
\beq
\Delta^\Phi_R=\frac{e^{2\Phi}}{\sqrt{{\rm det}(G)}} \partial_\mu
  e^{-2\Phi}\sqrt{{\rm det}(G)}G^{\mu\nu}\partial_\nu=\Delta_{R=1}+(R^2-1)\partial_\theta^2
+{(1-R^2)\over R^2}\partial_{\tilde{\theta}}^2\,,
\eeq
which shows that eigenfunctions of the Laplace operator corresponding to the Killing metric
on ${\rm SU}(2)$ which are also eigenfunctions of $\partial_\theta$, $\partial_{\tilde{\theta}}$
are in fact eigenfunctions of the generalized Laplacian $\Delta^\Phi$ for all $R$\footnote{Note that at $R=1$ $\Delta^\Phi=\Delta$ 
as discussed in the previous section.}, however with different
eigenvalues. For an eigenfunction of the Laplace operator in the irreducible ${\rm SU}(2)$ representation
labelled by $j\in\{0,\ldots,{k\over 2}\}$ with left and right ${\rm U}(1)$-quantum numbers $n$ and $\bar{n}$, the
difference of the eigenvalues of $\Delta^\Phi$ at $R$ and $R=1$ is given by
\beq\label{eigenvalues}
\delta^k_{j,n,m}(R) = \frac{R^2 -1}{4k} n^2 +\frac{1
  -R^2}{4kR^2} \bar{n}^2 ,
\eeq
where we have reintroduced the level $k$. 

Now, let us compare this with the CFT description \eq{deforb}.
The difference of ${1\over 2}\left(L_0+\bar{L}_0\right)$-eigenvalues
of a highest weight state in the $r$-twisted sector for $R$ and $R=1$
can be calculated to be
\begin{eqnarray}
d_{j,r,p}^k & = & 
 \frac{1}{4k} \left(
  \frac{1-R^2 }{R^2}m^2 + \left(
  R^2 
  -1\right) (m-r)^2\right)\,,\label{dimyang}
\end{eqnarray}
where we already identified the moduli spaces according to our general
discussion. 

We observe that (\ref{eigenvalues}) and
(\ref{dimyang}) agree, if we identify
\beq
n = m \,\,\, ,\,\,\, 
\bar{n} = m-r\,,
\eeq
and this identification is actually the one we expected from the
$\ZZ_k$-action \eq{Gammaaction}.
\section{Discussion}
Having shown that the effect of current-current deformations
of a conformal field theory on its structure is completely
captured by deformations of a charge lattice, we
obtained a description of the subspaces of CFT moduli spaces 
corresponding to these deformations as moduli spaces 
of certain lattices with additional structure. This generalizes
the case of deformations of toroidal conformal field theories
\cite{Dijkgraaf:jt}.

The general considerations were applied to WZW models, where
they were compared with a realization
of the deformed models as orbifolds of products of coset models
with varying toroidal models.
This realization was well suited for the construction
of sigma models corresponding to the deformed WZW models.
For this purpose we employed axial-vector
duality to transform the orbifold of the $\left(G/H \times H\right)$-sigma model
into a WZW-like model with (in general) non-bi-invariant metric, 
additional $B$-field and constant dilaton. This provides a very 
explicit description of the sigma models associated to deformed
WZW models. It would be interesting to investigate further the geometry of 
metric and $B$-field we obtained for the deformed models. Since the sigma models
correspond to conformal field theories, they should for example satisfy some nice
differential equations, namely the beta-function equations
(see \cite{Tseytlin:1989md} and references therein).

Apart from the general CFT considerations, we focused the discussion 
on the example of WZW models associated to compact, semi-simple Lie groups.
There are however other interesting conformal field theories
admitting current-current deformations, which deserve an analysis of their
moduli spaces. 
These are for example WZW models corresponding to non-semi-simple or 
non-compact Lie-groups (see e.g.\
\cite{Nappi:1993ie,Bars:1990rb,Forste:1994wp,Bars:1995mf,Manvelyan:2002xx}).
The former possess more complicated moduli spaces than semi-simple WZW models, 
because they give rise to more than one inequivalent deformation spaces,
while the latter
could provide time dependent exact string backgrounds and thus might 
give hints about 
how string theory deals with cosmological
singularities (see e.g.\
\cite{Liu:2002ft,Elitzur:2002rt,Elitzur:2002vw} for a discussion of
this point). It should be easy to modify the presented discussion in
such a way that one can obtain the models of Guadagnini et al.\ (see
e.g.\
\cite{Guadagnini:1987ty,Guadagnini:1987qc,PandoZayas:2000he,Quella:2002fk}
for discussions of those models) as 
limits in a class of 
deformed theories. One interesting aspect could be that these models
can be viewed as coset models with a trivial dilaton.   
Beyond a generalization to arbitrary WZW-models, 
also an investigation of moduli spaces of e.g.\ coset models, and
in particular Kazama-Suzuki-models 
\cite{Kazama:1988va} should be of interest, because the latter
would provide examples of 
explicitly known moduli spaces of $N=2$ superconformal field
theories. A discussion of 
the relation between mirror symmetry and gauge symmetry in this
setting has been presented in \cite{Giveon:1994cc}.

The analysis of the ``behaviour'' of D-branes (i.e.\ conformal boundary
conditions) on moduli spaces of conformal field theories was in fact 
the main motivation for this study of current-current deformations.
As our considerations show, these deformations are in fact easily tractable,
and hence provide a good setup to study ``bulk-deformations'' of boundary conformal
field theories.
Some semi-classical aspects of D-branes in deformed ${\rm SU}(2)$-WZW models
were presented in \cite{Forste:2001gn,Forste:2002uv}.
The general conformal field theory analysis of boundary conditions 
in deformed WZW models will be 
addressed in a forthcoming publication \cite{forthcoming}.
\subsubsection*{Acknowledgements}
The authors thank Matthias Gaberdiel, Werner Nahm, Jacek Pawelczyk,
Andreas Recknagel and Katrin Wendland for useful discussions. S.\ F.\
is grateful for the kind hospitality extended towards him during a
visit at Warsaw University where he had been given the opportunity to
present some of the results reported in this article. Especially, he
would like to thank Zygmunt Lalak.\\
D.\ R.\  was supported by DFG Schwerpunktprogramm 1096 and 
by the Marie Curie Training Site 
``Strings, Branes and Boundary Conformal Field Theory'' at King's 
College London, under EU grant HPMT-CT-2001-00296. 
The work of S.\ F.\  is supported in part by the European
Community's Human Potential 
Programme under contracts HPRN--CT--2000--00131 Quantum Spacetime,
HPRN--CT--2000--00148 Physics Across the Present Energy Frontier
and HPRN--CT--2000--00152 Supersymmetry and the Early Universe, and
INTAS 00-561.
\def\thesection{Appendix \Alph{section}}
\setcounter{section}{0}
\section{Deformation theory}
\label{deformationtheory}
\def\thesection{\Alph{section}}
In this appendix, techniques from conformal deformation theory (see e.g.\ 
\cite{Dijkgraaf:jt,Kutasov:1988xb,Ranganathan:1992nb,Ranganathan:1993vj})
are used to 
calculate  
the effect of current-current deformations on arbitrary conformal
field theories 
containing current algebras in their holomorphic and antiholomorphic
$W$-algebras. 

In the following, a family of conformal field theories is regarded 
as a Hermitian vector bundle over a differentiable
manifold\footnote{Singularities do not occur in our situation.}
parametrizing deformations of the conformal field theory structures in
a smooth way,  
i.e.\ all CFT structures are smooth sections of corresponding vector bundles.

Such families can be realized as perturbations \eq{deformation} by
exactly marginal fields. In this case, the tangent bundle of their
base manifolds 
are subbundles of  
the Hermitian vector bundles. The choice of regularization method and
renormalization scheme 
gives rise to connections on them \cite{Ranganathan:1993vj}. Here
connections $D$ 
(called $\ol{c}$ in \cite{Ranganathan:1993vj}) 
will be used, which restrict to the Levi-Civita connections on the
tangent bundles of the base  
manifolds equipped with the respective Zamolodchikov metrics. 
These connections are defined by ``minimal subtraction''
of divergences in the regularization constant.

Given a conformal field theory, it is 
in general quite hard to make 
global statements about the family of conformal field theories, 
generated by perturbation with a given set of exactly marginal
fields. This is due to the  
fact that information on the CFT structures in points of the family
have to be more or  
less completely reconstructed out of the structures at one point (the
point corresponding 
to the CFT which is being perturbed), by means of perturbation
theory. Thus, one gets perturbative results only, and perturbation
theory usually becomes technically difficult at higher orders.

However, in the case of perturbations by products of holomorphic and
antiholomorphic currents, there is in fact enough structure to make
exact global statements about the families of CFTs generated by them
using first order perturbation theory only. 

In \cite{Chaudhuri:1988qb} it was shown that tensor products of fields of
holomorphic and antiholomorphic currents are exactly marginal, if and
only if they form abelian current algebras  
$\hat{\ab}$, $\hat{\ol{\ab}}$ respectively, and that in this case 
the deformations generated by them preserve the corresponding current
algebras $\hat{\ab}$ and $\hat{\ol{\ab}}$. 
Thus these deformations give rise to families of conformal field theories with 
$\hat{\ab}$ and $\hat{\ol{\ab}}$ contained in their holomorphic and
antiholomorphic $W$-algebras. 
Moreover, the tangent vectors to the families in every point are given
by products of currents, 
whose CFT-properties are known independently of the actual CFT. Thus
the derivatives of the CFT-structures can be calculated in every point
of the families and can then be integrated up. 

Assuming that the Hilbert spaces of the conformal field theories in
the families decompose into $\hat{\ab}\times\hat{\ol{\ab}}$-highest weight
representations as in \eq{decomp} 
\beqn
\HHH\simeq \bigoplus_{(Q,\ol{Q})\in\Lambda}\HHH_{Q\ol{Q}}
\otimes\VV_Q\otimes\ol{\VV}_{\ol{Q}}\,.\nonumber 
\eeqn
it is shown in the following that these deformations only affect the
$\hat{\ab}\times\hat{\ol{\ab}}$- 
representations, while the OPE-coefficients of $\hat{\ab}\times\hat{\ol{\ab}}$-highest
weight states are
parallel with respect to the connection $D$. 
To be more precise, the only effect of the deformations will be 
${\rm O}(d,\ol{d})$-transformations of the charge lattices $\Lambda\in
\ab^*\times\ol{\ab}^*$.  From this it follows
in particular that the corresponding deformation spaces are given by
\eq{defspace} 
\beq\el{adefspace}
{\cal D}_{(a,\ol{a})}\simeq {\rm O}(d,\ol{d})_0/({\rm O}(d)\times {\rm O}(\ol{d}))_0
\simeq {\rm O}(d,\ol{d})/{\rm O}(d)\times {\rm O}(\ol{d})\,. 
\eeq
Let us first of all calculate the covariant derivatives of the modes
of the $\hat{\ab}$- and $\hat{\ol{\ab}}$-currents and holomorphic and
antiholomorphic energy-momentum tensors 
defined by
\beqn
j^\alpha(z)=\sum_{n\in\ZZ}z^{n-1}j_n^{\alpha}\,,
\qquad\ol{\jmath}^{\ol{\alpha}}(\ol{z})=
\sum_{n\in\ZZ}\ol{z}^{n-1}\ol{\jmath}_n^{\ol{\alpha}}\,,\\ 
T(z)=\sum_{n\in\ZZ}z^{n-2}L_n\,,\qquad
\ol{T}(\ol{z})=\sum_{n\in\ZZ}\ol{z}^{n-2}\ol{L}_n\,,\nonumber
\eeqn
where, as in section \ref{ccdef} $(j^\alpha)_\alpha$ and
$(\ol{\jmath}^{\ol{\alpha}})_{\ol{\alpha}}$ are basis of $\ab$ and
$\ol{\ab}$ respectively,  
and we denote the generators of the deformations by
${\cal O}^{\alpha\ol{\alpha}}(z,\ol{z}):=j^\alpha(z)
\ol{\jmath}^{\ol{\alpha}}(\ol{z})$.   
By the definition of $D$, $D_{{\cal O}^{\alpha\ol{\alpha}}}j_n^\beta$
can be expressed as 
\beq
D_{{\cal O}^{\alpha\ol{\alpha}}}j_n^\beta=\Big[\oint_{C(0)}{{\rm d}z
    \over 2\pi i}z^{-n} \int_{\CC P^1\backslash D_\epsilon(z)}{\rm
    d}^2w  O_{\alpha\ol{\alpha}}(w,\ol{w})j^\beta(z)\Big]_\epsilon\,,
\eeq
where $\epsilon$ is the regularization parameter and 
$[X]_\epsilon$ means the regularization independent part of $X$.
Using the OPE \eq{currentalg} this can be expressed as
\beqn
D_{{\cal O}^{\alpha\ol{\alpha}}}j_n^\beta&=&
\Big[\oint_{C(0)}{{\rm d}z\over 2\pi i}z^{-n}
  \oint_{C_\epsilon(z)}{{\rm d}\ol{w}\over -2i}  
{kK^{\alpha\beta}\ol{\jmath}^{\ol{\alpha}}(\ol{w})\over (z-w)}\Big]_\epsilon\\
&=&
\Big[\oint_{C_\epsilon(0)}{{\rm d}\ol{w}\over 2i} 
{kK^{\alpha\beta}\ol{\jmath}^{\ol{\alpha}}(\ol{w})}
\oint_{C(w)}{{\rm d}z\over 2\pi i}{z^{-n}\over (z-w)}
\Big]_\epsilon\nonumber\\
&=&
\Big[\oint_{C_\epsilon(0)}{{\rm d}\ol{w}\over 2i} 
{kK^{\alpha\beta}\ol{\jmath}^{\ol{\alpha}}(\ol{w})}w^{-n}
\Big]_\epsilon\nonumber\\
&=&
-\pi k K^{\alpha\beta}\ol{\jmath}_0^{\ol{\alpha}}\delta_{n,0}\,.\nonumber
\eeqn
The same kind of arguments lead to 
\beq
D_{{\cal O}^{\alpha\ol{\alpha}}}L_n^\beta=
\Big[\oint_{C(0)}{{\rm d}z\over 2\pi i}z^{-n} \int_{\CC P^1\backslash
    D_\epsilon(z)}{\rm d}^2w  
O_{\alpha\ol{\alpha}}(w,\ol{w})T(z)\Big]_\epsilon
=-\pi j_n^\alpha\ol{\jmath}_0^{\ol{\alpha}}
\eeq
and similar expressions for the modes of $\hat{\ol{\ab}}$. Altogether we find
\beqn\el{dermodes}
D_{{\cal O}^{\alpha\ol{\alpha}}}j^\beta_n=
-\pi k K^{\alpha\beta}\delta_{n,0}\ol{\jmath}^{\ol{\alpha}}_n\,,\qquad
D_{{\cal O}^{\alpha\ol{\alpha}}}L_n=
-\pi j_n^\alpha\ol{\jmath}_0^{\ol{\alpha}}\,,\\
D_{{\cal O}^{\alpha\ol{\alpha}}}\ol{\jmath}^{\ol{\beta}}_n=
-\pi k\ol{K}^{\ol{\alpha}\ol{\beta}}
\delta_{n,0}j^\alpha_n\,,\qquad
D_{{\cal O}^{\alpha\ol{\alpha}}}\ol{L}_n=
-\pi j_0^\alpha\ol{\jmath}_n^{\ol{\alpha}}\,.\nonumber
\eeqn
Thus, the $\hat{\ab}\times\hat{\ol{\ab}}$-$W$-algebra structure is
parallel with respect  
to $D$, and only the $\ab\times\ol{\ab}$-charges\footnote{As noted above,
  we assume the zero modes of the  
currents to be diagonalizable on $\HHH$.} change under the deformations.
Their covariant derivatives can be read off from \eq{dermodes}
\beq\el{defcharges}
D_{c_{\alpha\ol{\alpha}}{\cal O}^{\alpha\ol{\alpha}}}(Q^\beta,
\ol{Q}^{\ol{\beta}})= 
-\pi (c_\alpha^{\ol{\beta}}Q^\alpha,c^\beta_{\ol{\alpha}}
\ol{Q}^{\ol{\alpha}})\,, 
\eeq
which are transformations  
in $o(\ab^*\oplus \ol{\ab}^*,\kappa-\ol{\kappa})$. This characterizes
completely the deformations of the
$\hat{\ab}\times\hat{\ol{\ab}}$-$W$-algebra structures of the CFTs. 
Moreover the $\hat{\ab}\times\hat{\ol{\ab}}$-highest weight property and
the decomposition of the Hilbert space \eq{decomp} are preserved under
the deformations, which means in  
particular that we only have to assume \eq{decomp} for one CFT in the
family. 

To show that this is in fact the only effect of the deformations on the
CFT structures, 
we have to show, that the OPE of $\hat{\ab}\times\hat{\ol{\ab}}$-highest
weight vectors is not deformed. Now, the covariant derivative of
correlation functions of those vectors is given by
\beqn\el{defope}
&&D_{{\cal O}^{\alpha\ol{\alpha}}}\bra\Phi_1(z_1,\ol{z}_1)
\ldots\Phi_n(z_n,\ol{z}_n)\ket\\ 
&&\qquad
=\Big[\int_{\CC P^1\backslash\bigcup_i D_\epsilon(z_i)} {\rm d}^2w
  \bra{\cal O}^{\alpha\ol{\alpha}} 
\Phi_1(z_1,\ol{z}_1)\ldots\Phi_n(z_n,\ol{z}_n)\ket\Big]_\epsilon\nonumber\\
&&\qquad
=\sum_{i,j}Q_i^\alpha \ol{Q}_j^{\ol{\alpha}}
\bra\Phi_1(z_1,\ol{z}_1)\ldots\Phi_n(z_n,\ol{z}_n)\ket
\Big[\int_{\CC P^1\backslash\bigcup_i D_\epsilon(z_i)}{-1\over
    (w-z_i)(\ol{w}-\ol{z}_j)}\Big]_\epsilon 
\nonumber\\
&&\qquad
=-2\pi\sum_{i<j}Q_i^\alpha\ol{Q}_j^{\ol{\alpha}}ln\left(|z_i-z_j|^2\right)
\bra\Phi_1(z_1,\ol{z}_1)\ldots\Phi_n(z_n,\ol{z}_n)\ket\nonumber\\
&&\qquad
=\left(D_{{\cal O}^{\alpha\ol{\alpha}}} 
ln\left(\prod_{i<j}
(z_i-z_j)^{2\kappa(Q_i,Q_j)}(\ol{z}_i-\ol{z}_j)^{2\ol{\kappa}(\ol{Q}_i,
  \ol{Q}_j)}\right)\right)
\nonumber\\
&&\qquad\qquad\qquad\qquad\qquad\qquad\qquad\qquad\qquad
\times\bra\Phi_1(z_1,\ol{z}_1)\ldots\Phi_n(z_n,\ol{z}_n)\ket\,.\nonumber
\eeqn
But the $ln$-term in the last line of \eq{defope} is just the logarithm of
the corresponding $\hat{\ab}$-, $\hat{\ol{\ab}}$-conformal block. 
Thus, the correlation functions are deformed only through the 
conformal blocks and the OPE-coefficients of
$\hat{\ab}\times\hat{\ol{\ab}}$-highest weight states 
are parallel.

Thus, the effect of current-current deformations on the conformal
field theory structure is completely characterized by the deformations
of the charges described above.  
In particular from \eq{defcharges} it follows that the base manifold
of the family of CFTs  
generated by current-current deformations is indeed given by \eq{adefspace}. 

Let us finish with a comment on another connection $\tilde{D}$.
In the discussion above, we used connections $D$ on the Hilbert space
bundles over the deformation spaces, which were defined by minimal
subtraction. With respect to these connections operators from the
$W$-algebras are orthogonal to their zero mode of the current algebra.
For the discussion of e.g.\ boundary conditions other connections
$\tilde{D}$ will be useful. These are defined by\footnote{In fact they
  are nothing else  
than the connections $\hat{\Gamma}$ from \cite{Ranganathan:1993vj},
which are obtained by a regularization scheme consisting of cutting
out radius one  
disks around the punctures of the surfaces.}
\beq
\tilde{D}_{{\cal O}_{\alpha\ol{\alpha}}}:=D_{{\cal O}_{\alpha\ol{\alpha}}}-
\pi\sum_{n\neq 0}{1\over n}j_n^\alpha\ol{\jmath}_n^{\ol{\alpha}}\,.
\eeq
They satisfy
\beqn\el{tildedermodes}
\tilde{D}_{{\cal O}^{\alpha\ol{\alpha}}}j^\beta_n=
-\pi k K^{\alpha\beta}\ol{\jmath}^{\ol{\alpha}}_{-n}\,,\qquad
\tilde{D}_{{\cal O}^{\alpha\ol{\alpha}}}L_n=
-\pi \sum_m j_{m+n}^\alpha\ol{\jmath}_m^{\ol{\alpha}}\,,\\
\tilde{D}_{{\cal O}^{\alpha\ol{\alpha}}}\ol{\jmath}^{\ol{\beta}}_n=
-\pi k\ol{K}^{\ol{\alpha}\ol{\beta}}
j^\alpha_{-n}\,,\qquad
\tilde{D}_{{\cal O}^{\alpha\ol{\alpha}}}\ol{L}_n=
-\pi \sum_m j_m^\alpha\ol{\jmath}_{m+n}^{\ol{\alpha}}\,,\nonumber
\eeqn
and thus for all $n\in\mathbb{Z}$, the parallel transport of
$(j_n^\alpha,\ol{\jmath}_{-n}^{\ol{\alpha}})_{\alpha,\ol{\alpha}}$ 
is given by the vector representation of ${\rm O}(d,\ol{d})$.\\
As the connections $D$, also $\tilde{D}$ restrict to the
Levi-Civita-connection on the 
tangent bundle of the deformation space equipped with the Zamolodchikov metric.
%
%

%
%
\providecommand{\bysame}{\leavevmode\hbox to3em{\hrulefill}\thinspace}

\end{document}